\documentclass[preprint,showpacs,preprintnumbers,amsmath,amssymb]{revtex4}

\usepackage{graphicx}
\usepackage{dcolumn}
\usepackage{bm}

\def\D{\partial}
\def\grad{\nabla}
\def\div{\nabla \cdot}
\def\rot{\nabla \times}
\def\lap{\nabla^2}
\def\inv{^{-1}}

\def\Eq#1{(\ref{eq:#1})}
\def\EEq#1{Eq.(\ref{eq:#1})}
\def\EEEq#1{Equation (\ref{eq:#1})}
\def\Tab#1{Table \ref{tab:#1}}   
\def\Fig#1{Fig.\ref{fig:#1}}   

\def\eq{\begin{eqnarray}}
\def\qe{\end{eqnarray}}

\def\eqnn{\begin{eqnarray*}}
\def\qenn{\end{eqnarray*}}

\def\seq{\begin{subequations}\begin{eqnarray}}
\def\sqe{\end{eqnarray}\end{subequations}}

\def\nn{\nonumber}
\def\uprime{^{'}}
\def\deg{^\circ}

\def\bD{\bm{D}}

\def\bM{\bm{M}}
\def\bN{\bm{N}}

\def\bR{\bm{R}}

\def\bc{\bm{c}}

\def\be{\bm{e}}

\def\bem{\bm{m}}
\def\bn{\bm{n}}

\def\bq{\bm{q}}
\def\br{\bm{r}}

\def\bv{\bm{v}}

\def\simge{\;\lower3pt\hbox{$\stackrel{\textstyle >}{\sim}$}\;}
\def\simle{\;\lower3pt\hbox{$\stackrel{\textstyle <}{\sim}$}\;}

\def\para{\parallel}

\def\t#1{\tilde{#1}}
\def\bm#1{\mbox{\boldmath $#1$}}

\def\lrL#1{\left[#1\right]}
\def\lrM#1{\left\{#1\right\}}
\def\lrS#1{\left(#1\right)}
\def\lrF#1{\left|#1\right|}
\def\lrA#1{\left\langle #1 \right\rangle}

\def\mycomment#1{}

\def\f#1#2{\frac{#1}{#2}}
\def\der#1#2{\f{\D #1}{\D #2}}
\def\fder#1#2{\f{\delta #1}{\delta #2}}

\def\Cpl{\t{C}_{\para}}
\def\Cpr{\t{C}_{\perp}}
\def\Dpd{\t{D}^{'}}
\def\Dpl{\t{D}_{\para}}
\def\Dpr{\t{D}_{\perp}}
\def\Dpp{\t{D}_{\para\perp}}
\def\Np{N_{\perp}}
\def\bqp{\bq_{\perp}}
\def\qpr#1{\t{q}_{\perp#1}}
\def\qp{q_{\perp}}
\def\kp{k_{\perp}}
\def\talpha{\t{\alpha}}
\def\tk1{\t{\kappa}_1}
\def\k2{\t{\kappa}_2}
\def\tc{\t{\bc}}
\def\tF{\t F}

\begin{document}

\preprint{APS/123-QED}

\title{Effective elastic theory of smectic-A and smectic-C liquid crystals}

\author{Hiroto Ogawa}
 \email{hiroto@cmpt.phys.tohoku.ac.jp}
\affiliation{%
Department of Physics, Tohoku University, Sendai, 980-8578, Japan\\
}%

\date{\today}

\begin{abstract}
We analytically derive the
effective layer elastic energy of smectic-A and smectic-C liquid crystals
by
adiabatic elimination of the orientational degree of freedom
from the generalized Chen-Lubensky model.
In the smectic-A phase,
the effective layer bending elastic modulus is calculated as a function of the
wavelength of the layer undulation mode.
It turns out that an unlocking of the layer normal and the director
reduces the layer bending rigidity
for wavelengths smaller
than the director penetration length.
In the achiral smectic-C phase, an anisotropic bending
elasticity appears due to the coupling between
the layer displacement and director.
The effective layer bending rigidity is calculated as a function of the angle $\vartheta$
between the layer undulation wave-vector and the director field.
We compute the free energy minimizer
$\vartheta=\theta$.
It turns out that $\theta$ varies from $0\deg$ to $90\deg$ depending on the tilt angle, undulation wave-length and other elastic constants.
We also discover a new important characteristic length and the discontinuous change of $\theta$.
Using the elastic constants of Chen-Lubensky model,
we determine the parameters of the more macroscopic model [Y. Hatwalne and T. C. Lubensky, Phys. Rev. E {\bf 52}, 6240 (1995)].
We then discuss the hydrodynamics,
and demonstrate the alignment of director and the propagation of the anisotropic layer displacement wave in the presence of
an oscillatory wall and a vibrating cylindrical source respectively.
\end{abstract}

\maketitle

\section{\label{sec:level1}INTRODUCTION}
\par
Liquid crystals have many fascinating physical properties,
because they have both the molecular position and the molecular orientation as degrees of freedom~\cite{deGennes}.
This multiplicity gives rise to
interesting phase transition phenomena,
including the tricritical and the Lifshitz point in the phase diagram
~\cite{Lubensky},
where
the first and second order transitions are switched~\cite{Keyes, Aharony},
and three phases (i.e., the
nematic, smectic-A (Sm-A) and smectic-C (Sm-C)) coexist~\cite{Shashidhar, Drossinos}, respectively.
To explain the phase behavior around the nematic-Sm-A-Sm-C (NAC) Lifshitz point theoretically,
a variety of models have been developed~\cite{deGennesSmC, Chu, ChenLubensky, Benguigui, Huang, Grinstein}.
De Gennes~\cite{deGennesSmC} and Chu and McMillan~\cite{Chu} introduced a tilt angle and an in-plane director as new degrees of freedom, respectively.
Chen and Lubensky~\cite{ChenLubensky} took fourth order derivative of the density field into account.
Benguigui~\cite{Benguigui} and Huang and Lien~\cite{Huang} adopted a smectic-C scalar order parameter together with a smectic-A order parameter.
Grinstein and Toner~\cite{Grinstein} also used the two-dimensional in-plane director component to describe the free energy.
Among them, the Chen-Lubensky model is supported by quite a few X-ray scattering experiments
~\cite{Witanachchi, Safinya, Martinez}.
%
\par
In the smectic phase,
the
layer order and the director often cause a frustration.
In chiral liquid crystals, geometrical incompatibility of
a uniform smectic order and a helical director configuration
leads to
a rich variety of defect phases~\cite{Kitzerow}.
They can be discussed with the Landau-de Gennes model,
utilizing an analogy with superconductors in the magnetic field,
where the smectic order parameter, the director vector and the chirality
are identified with the wave function of a superconducting particle,
the electromagnetic vector potential
and the external magnetic field, respectively~\cite{deGennesAna}.
The twist-grain-boundary (TGB) phase is the simplest defect phase
where the groups of planes, containing parallel screw dislocations, are regularly stacked
with the dislocations in adjacent planes tilted each other at a constant angle,
and finite length smectic slabs are inserted between the dislocation planes~\cite{Renn, Goodby}.
The Chen-Lubensky model, a higher order extension of the Landau-de Gennes model,
is again successful in explaining the structure and phase transition of the TGB$_{\text{A}}$ and TGB$_{\text{C}}$ phases,
which have the Sm-A and Sm-C slabs respectively~\cite{LubenskyRenn, Ismaili}.
More complex defect structures such as cholesteric and smectic blue phases have been discovered~\cite{Kitzerow}.
The cubic smectic blue phase (Sm-BP$_{\text{I}}$) is comprised of a three-dimensional cubic disclination lattice,
whose lattice constant is in the order of the wavelength of the visible light~\cite{Grelet, DiDonna}.
The isotropic smectic blue phase (Sm-BP$_{\text{Iso}}$) has fluid-like rheological properties and the detailed structure is completely unknown~\cite{Yamamoto}.
Such novel defect structures
in principle, should be understood with the Landau-de Gennes and Chen-Lubensky models presented above.
However, in fact, these phenomenological models are not appropriate for theoretical explanations of these phases,
because of the highly complex spatial structure together with the quite intricate free energy model.
DiDonna and Kamien make use of a simpler phenomenological model to discuss the stability of the cubic blue phase~\cite{DiDonna}.
In their Landau-Peierls form, the director is assumed to be parallel to the layer normal.
However, a tilt of the director from the layer normal is reported
to weaken the layer bending elasticity in the numerical simulation of the TGB$_{\text{A}}$ phase~\cite{OgawaTGB},
and to change the structural symmetry of the smectic blue phases in the experimental study~\cite{GreletTilt}.
Such anomalous properties in the TGB and the smectic blue phases
have their origin in a nontrivial elastic mechanism of the undulated smectic layer,
combined with the director degree of freedom.
Therefore, a modification of the Landau-Peierls model,
accommodating the director elasticity of Sm-A and Sm-C phases,
would help the understanding of the complex layered structures.
In addition, the Landau-Peierls model has the same layer compression and bending elastic energy forms
as those of the other layered materials such as block copolymers and surfactant system~\cite{OhtaKawasaki, GompperKlein}.
Thus, the characteristic features of liquid crystals are not very apparent in the original Landau-Peierls model.
The director in liquid crystals should affect the layer elasticity,
while block copolymers do not possess an orientational degree of freedom especially in the weak segregation limit~\cite{Leibler}.
Thus it is worthwhile to derive an intermediately simple model
directly from the Chen-Lubensky free energy,
eliminating the variation of the director degree of freedom,
to clarify the role of the director elasticity in smectic liquid crystals.
At the same time, by doing so,
we can obtain the phenomenological elastic constants of the macroscopic models~\cite{DiDonna, Hatwalne},
in terms of the more microscopic parameters of Chen-Lubensky model.
\par
Dynamics of Sm-C layers is also an interesting topic of liquid crystals~\cite{deGennes, MartinParodi, Buka, Pargellis, Carlsson}.
The director component parallel to layer ($\bc$-vector) plays
an important role on the static and dynamic pattern formation due to the anisotropic elasticity and interaction with flow field~\cite{Johnson, Cladis}.
Shear flow orients the $\bc$-vector and results in a novel target pattern with a disclination~\cite{Cladis, Chevallard}.
On the other hand, because of the static coupling between the layer displacement and the director,
together with the rotational viscosity,
an oscillatory wave traveling perpendicularly to the layer normal also rotates and aligns the $\bc$-vector in a certain direction,
even in a uniform oscillatory wave~\cite{deGennes, ClarkSmC}.
It might be useful for an application, for instance, mechanically-active optical devices
and sensitive acoustic sensors~\cite{ClarkSmA, ClarkSmC, Yablonskii, Uto}.
\par
This paper is organized as follows.
In Section II, we review the original Landau-de Gennes model
and derive the effective elastic energy of Sm-A phase.
The analysis is extended to Sm-A and Sm-C phases
with the generalized Chen-Lubensky model in Section III.
Next we conduct a hydrodynamic simulation for Sm-C layer in Section IV.
We conclude in Section V.

\section{\label{sec:level1}Landau-de Gennes model for Smectic-A phase}

In this section, we briefly review the Landau-de Gennes model for Sm-A phase,
and then calculate the effective layer elastic energy in terms of the layer displacement field.
The physical meaning of the results are discussed qualitatively.
\par
Smectic liquid crystals can be described by the density modulation $\Psi(\br)$
and the director $\bn(\br)$.
The density field $\Psi(\br)$ is a complex order parameter
with the absolute value being the amplitude of the smectic order
and the phase describing the layer displacement~\cite{deGennesAna}.
The analogy between liquid crystals and superconductors leads to
the phenomenological Landau-de Gennes model for the Sm-A liquid crystals
\seq
F&=&F_{\text{layer}} + F_{\text{cpl}} + F_{\text{Frank}},
\label{eq:L-deG}\\
F_{\text{layer}}&=&\int d\br \lrL{\tau|\Psi|^2+\f{g}2|\Psi|^4},\\
F_{\text{cpl}}&=&\int d\br \f{1}{2} \left[ B_{\para} \lrF{\lrS{\bn\cdot\grad-iq_0}\Psi}^2 +B_{\perp}\lrF{\bn\times\grad \Psi}^2 \right],\nn\\
\label{eq:L-deGcpl}\\
F_{\text{Frank}}&=&\int d\br \f{1}{2}\left[K_1(\div{\bn})^2+K_2(\bn\cdot\rot{\bn})^2\right.\nn\\
&&\left.+K_3(\bn\times\rot{\bn})^2\right].
\sqe
The double-well potential $F_{\text{layer}}$ describes the Nematic-Sm-A (NA) transition.
$F_{\text{cpl}}$ and $F_{\text{Frank}}$ are the coupling energy between $\Psi$ and $\bn$,
and the Frank elastic energy.
The dimensionless temperature $\tau$ is positive (negative) above (below) the NA point.
The constants $B_{\para}$ and $B_{\perp}$ are the layer compression elastic coefficients which adjust the layer width to the equilibrium value $d=2\pi/q_0$.
The Frank elastic constants $K_i \, (i=1, 2, 3)$ correspond to
the splay ($i=1$), the twist ($i=2$) and the bend ($i=3$) deformations, respectively.
\par
\begin{figure}[htbp]
 \leavevmode
 \begin{center}
  \begin{tabular}{lclcl}
	&&(a) $\qp \ll \lambda\inv$
	&& (b) $\qp \agt \lambda\inv$\\
	\includegraphics[width=15mm]{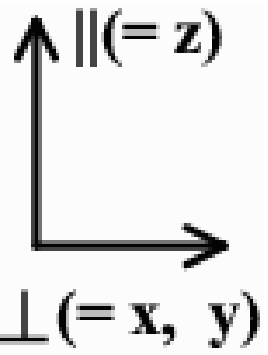}
	&&\includegraphics[width=27mm]{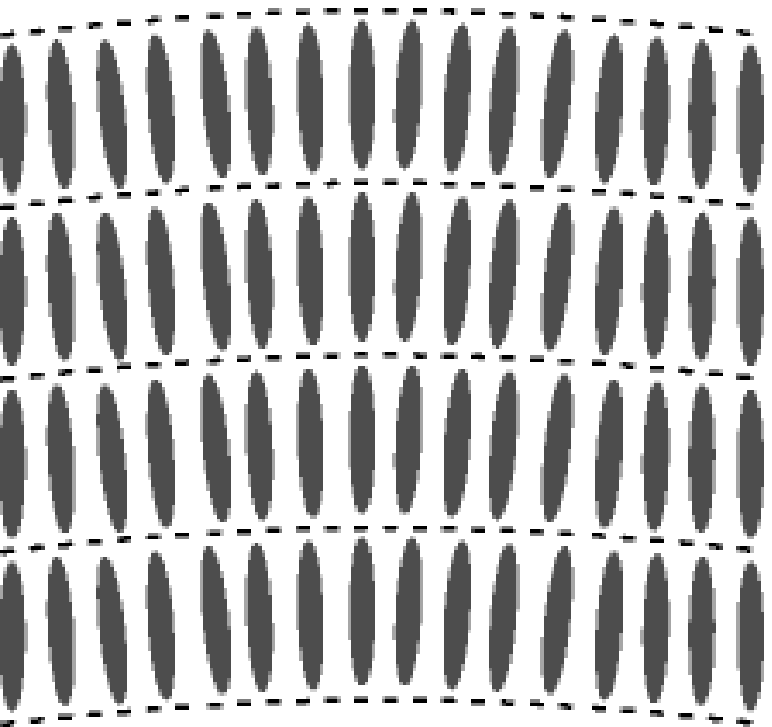}
	&&\includegraphics[width=27mm]{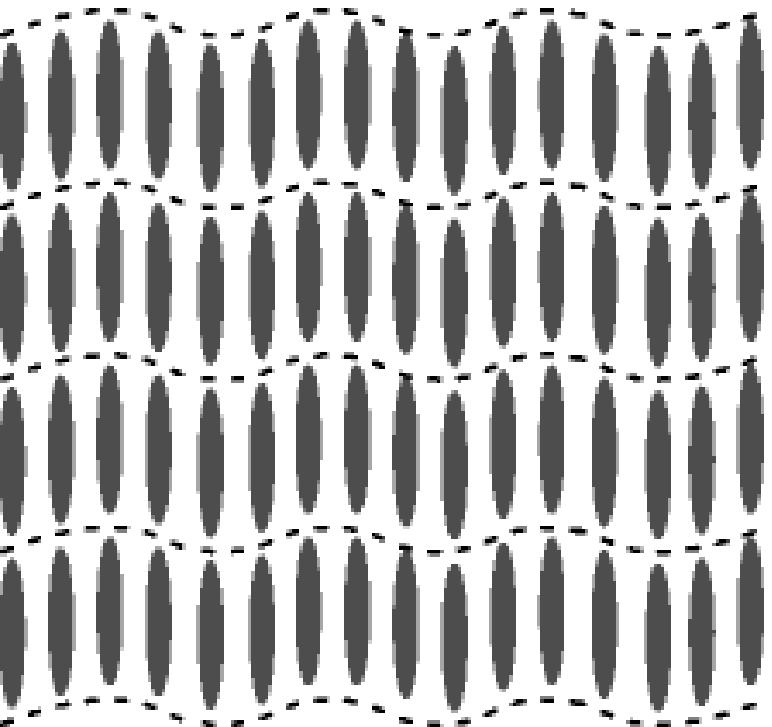}
  \end{tabular}
  \caption{Schematic representation of the undulated layer structure of Sm-A phase.
The director and layer normal are (a) almost parallel for $\qp \ll \lambda\inv$
	and (b) decoupled for $\qp \agt \lambda\inv$.}
  \label{fig:SmAq}
 \end{center}
\end{figure}

We consider a perturbation of the uniform Sm-A structure far below the NA point
to see the effect of the director tilt from the layer normal.
For this purpose, we set $B_{\para}=B_{\perp}\equiv B$ and $K_1=K_2=K_3 \equiv K$ to simplify the discussion.
A more general case is considered in the next section for the Chen-Lubensky model.
The amplitude of the smectic order is close to the equilibrium value at low temperature, so we assume
\eq
\Psi(\br)=\Psi_0\exp[iq_0(z-u(\br))]\,\,\,\lrS{\Psi_0=\sqrt{\f{\tau}{g}}},
\qe
where the $z$-axis is set along the equilibrium layer normal,
and the layer displacement field $u(\br)$ is introduced.
The director $\bn$ can be divided into its spatial average and the deviation:
\eq
\bn(\br)=\be_z+\delta\bn(\br).
\qe
The free energy components $F_{\text{cpl}}$ and $F_{\text{Frank}}$ are now simplified as
\seq
F_{\text{cpl}}&=&\f{K}{2\lambda^2} \int d\br \lrF{\grad u+\delta\bn}^2,
\label{eq:L-deGc}
\\
F_{\text{Frank}}&=&\f{K}{2} \int d\br \lrF{\grad\delta\bn}^2,
\label{eq:L-deGf}
\sqe
where $\lambda=\sqrt{K/B}/(\Psi_0q_0)$ is the penetration length~\cite{Renn},
and $\lrF{\grad\delta\bn}^2=(\D_i\delta n_j)(\D_i\delta n_j)$ (repeated indices are summed up).
We next write the linearized effective free energy
in terms of $u(\br)$.
To do this, the director is adiabatically eliminated with the equation,
\eq
(\b1-\bn\bn)\cdot\fder{F}{\bn}=0.
\label{eq:normal}
\qe
The factor $\b1-\bn\bn$ ensures the normalization $\lrF{\bn}^2=1$,
and $\delta n_z=0$ because $\lrF{\be_z+\delta\bn}=1$.
Using \Eq{L-deGc}, \Eq{L-deGf} and \Eq{normal}, 
we obtain the Fourier representations
\eq
\delta n_j(\bq)&=&-\tilde{\kappa}(q) iq_j u(\bq).
\label{eq:L-deGnu}\\
\biggl( \tilde{\kappa}(q) &\equiv& \f{1}{1+\lambda^2q^2}, \,\,\, j=1, 2 \biggr).\nn
\qe
Thus the effective free energy is
\eq
F_{\text{eff}}&=&\f{K}{2}\int_{\bq} \lrL{ \lrS{ \f{q_z}{\lambda} }^2 + \tilde{\kappa}(q)q_{\perp}^4 + \tilde{\kappa}(q)q_{\perp}^2q_z^2} \lrF{u(\bq)}^2,
\label{eq:effL-deG}
\qe
where we define the $z$-axis and in-plane (x- and y-) axes as the parallel and perpendicular direction respectively (\Fig{SmAq}).
\EEEq{effL-deG} is identical with the Landau-Peierls free energy
except for
the Lorentzian dependence of the dimensionless layer bending elastic coefficient $\tilde{\kappa}(q)$,
and the higher order cross term~\cite{Lubensky, ClarkSmA}.
%
\par
The $q$-dependence of $\tilde\kappa(q)$
is interpreted qualitatively as follows.
We consider a pure undulation case with a constant layer thickness $q_{\parallel}=0$.
This condition is, in fact, a sufficiently good approximation to describe experiments in the thermodynamic limit~\cite{ClarkSmA}.
Thus $\bq \parallel \delta\bn$ is satisfied,
which means that only the splay part $\lrF{\bq\cdot\delta\bn}^2$ is nonzero in the three Frank elastic contributions.
Since the Landau-de Gennes model \Eq{L-deG} has no layer bending term in the $\Psi$-dependent part,
the effective elasticity comes solely from the splay term.
In fact, if the director $\bn$ coincides with the layer normal $\bem$,
the splay term is turned into the layer bending elastic energy:
\eq
\f{K}{2}\int d\br \lrS{\div\bn}^2=2K\int d\br H^2.
\label{eq:splayH}
\qe
Here we denote the mean curvature of the smectic layer by $H$.

\EEEq{splayH} is achieved well for $\qp\inv \gg \lambda$,
where the director is locked to the layer normal (\Fig{SmAq}(a)).
Oppositely, for $\qp\inv \alt \lambda$,
the director cannot follow the layer deformation,
and the splay term less contributes to the layer elasticity (\Fig{SmAq}(b)).
We note that $\tilde\kappa(q)$ can be written also as a function of the root mean squared (RMS) tilt angle $\bar{\alpha}=\sqrt{\lrA{\alpha^2}}$.
With a smooth undulation $u(\br)=u_0\cos\lrS{\bqp\cdot\br}\,\,(\qp \ll \lambda\inv)$,
\eq
\tilde{\kappa}(\bar{\alpha})=\f{1}{1+\lrS{\sqrt{2}\bar{\alpha}\lambda/u_0}^{2/3}}.
\label{eq:Keffa}
\qe
The detailed derivation is given in Appendix A.
The effective layer bending rigidity decays with $\bar{\alpha}$. It implies that the
coupling between $\bem$ and $\bn$ is certainly essential.

In fact, for any $\Psi$, the locking term is found in the coupling term \Eq{L-deGcpl}
\begin{widetext}
\eq
F_{cpl}=\f{1}{2}\int d\br \left[ B_{\para} \biggl| \bn \cdot \biggl( - i\grad{\Psi_0}
+ \Psi_0\bn(|\grad{\phi}|-q_0) + \Psi_0|\grad{\phi}|(\bem-\bn) \biggr) \biggr| ^2 \right.\nn\\
\left. + B_{\perp} \biggl| \bn \times \biggl( - i\grad{\Psi_0}
+ \Psi_0|\grad{\phi}|\bem \biggr) \biggr| ^2\right]\nn\\
\label{eq:L-deGdecomp}
\qe
\end{widetext}
where the complex order parameter is decomposed into the amplitude and the phase: $\Psi(\br)=\Psi_0(\br)\exp(i\phi(\br))$,
and the gradient of the phase component defines the layer normal vector $\bem(\br)=\grad{\phi}/|\grad{\phi}|$.
In the parallel part, the first term favors a uniform smectic modulation amplitude, the second is the layer compression energy
and the last term locks the director along the layer normal.
In the perpendicular part, on the other hand, the first term is again the amplitude homogenizing contribution,
and the second term reduces the director component perpendicular to the layer normal
($\bc$-vector) as long as the coefficient $B_{\perp}$ is positive.
The Sm-C phase is stable with a negative $B_{\perp}$,
as we shall see in the next section.
\section{\label{sec:level1}Generalized Chen-Lubensky model for smectic-A and -C phases}
In this section, we introduce the Chen-Lubensky model in the generalized way.
Then the effective elastic energy is calculated for both Sm-A and Sm-C phases.
After a physical interpretation of the Sm-A energy,
the effective layer bending elasticity of the Sm-C phase is discussed.
The anisotropy is characterized by the angle $\vartheta$ between the $\bc$-vector and the layer bending direction.
The state diagrams for the easiest bending angle $\theta$ are studied.
The free energy as a function of $\vartheta$ is also calculated.
Finally, the model parameters of the more macroscopic free energy~\cite{Hatwalne} are
determined with those of Chen-Lubensky model.
\subsection{Generalized Chen-Lubensky model}
By adding higher order gradient terms,
one can extend the Landau-de Gennes free energy
to reproduce the achiral Sm-C phase.
The Chen-Lubensky model is given in the most generalized form:
\seq
F&=&\f{K_3}{2}\tF\,\,\,\,\,\,\,\nn\\
\left( \tF \right. &=& \left.\tF_{\text{layer}}+\tilde{F}_{\text{cpl}}^{(2)}+\tilde{F}_{\text{cpl}}^{(4)}+\tilde{F}_{Frank} \right) ,
\label{eq:CL}\\
\Psi_0^2 q_0^2 \tF_{\text{layer}}&=& 2\int d\br \lrL{ \t{\tau} |\Psi|^2+\f{ \t{g} }2|\Psi|^4},\\
\Psi_0^2 q_0^2 \tF_{\text{cpl}}^{(2)}&=& 2\int d\br \left[\lrS{\Cpl n_i n_j + \Cpr \delta_{ij}^{\perp}} (D_i\Psi) (D_j\Psi)^{*} \right],\nn\\
\Psi_0^2 q_0^2\tF_{\text{cpl}}^{(4)}&=& 2\int d\br \left[ \Dpl \lrF{n_i n_j D_i D_j\Psi}^2 \right.\nn\\
&&+ \Dpp \lrS{(n_i n_j D_i D_j\Psi)\lrS{\delta_{kl}^{\perp}D_k D_l\Psi}^{*}+c.c.}\nn\\
&&\left. + \Dpr \lrF{\delta_{ij}^{\perp} D_i D_j\Psi}^2 \right],\\
\tF_{\text{Frank}}&=&\int d\br \left[ \lrS{1+\tk1}(\div{\bn})^2 \right.\nn\\
&& + \lrS{1+\k2} K_2(\bn\cdot\rot{\bn})^2 \nn\\
&&\left.+\lrM{\bn\times\lrS{\rot{\bn}}}^2\right],\nn\\
\sqe
where $D_i=\D_i-iq_0 n_i$ is the covariant derivative and $\delta_{ij}^{\perp}=\delta_{ij}-n_i n_j$
is the projection operator.
The elastic constants are $\Cpl$ and $\Cpr$ for the second order gradient term $\tF_{\text{cpl}}^{(2)}$,
$\Dpl$, $\Dpp$ and $\Dpr$ for the fourth order terms $\tF_{\text{cpl}}^{(4)}$.
%
\begin{figure}[htbp]
 \includegraphics[width=45mm]{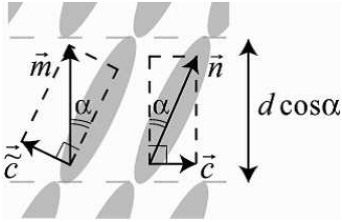}
 \caption{Schematic representation of the $\bc$ and $\t{\bc}$ vector
defined by the layer normal vector $\bem$ and the director $\bn$.
It is obvious that $\t{\bc} \rightarrow -\bc$ in the small tilt limit $\alpha \rightarrow 0$.}
 \label{fig:cov}
\end{figure}
The physical meaning of the covariant derivative is as follows (\Fig{cov}).
With the sinusoidal density profile $\Psi(\br)=\Psi_0 \exp\lrS{i\phi(\br)}$
and the layer normal vector $\bem \equiv \grad\phi / |\grad\phi|$,
one obtains $\bD\Psi=i|\grad\phi|(\bem-\bn\cos\alpha)\Psi \equiv i\t{\bc}|\grad\phi|\Psi$,
where $\alpha$ is the angle between $\bn$ and $\bem$, and $\tc$ is the layer normal vector component perpendicular to the molecular orientation,
which coincides with $-\bc$ for $\alpha \rightarrow 0$.
%
Because $\Cpl|\bn\cdot\bD\Psi|^2=0$ and $\Cpr|\bn\times\bD\Psi|^2=\Cpr\Psi_0^2|\grad\phi|^2\tc^2$,
Sm-C phase is equilibrated with a combination of a negative $\Cpr$
and stabilizing fourth-order covariant derivatives.
\par
The scattering function is readily calculated as
\begin{widetext}
\eq
S(\bq)=\f{S_0\t{\tau}_0}{ \t{\tau}_0+\Dpl(q_{\para}-q_0)^4+(\Cpl+2\Dpp q_{\perp}^2)(q_{\para}-q_0)^2+\Dpr (q_{\perp}^2-q_{0\perp}^2)^2 },
\qe
\end{widetext}
where
we set $\t{\tau}_0=\t{\tau}-\Cpr^2/(4\Dpr)$ and $q_{0\perp}^2=-\Cpr/(2\Dpr)$.
The relations of the present generalized model to the other models are shown in \Tab{CL}.
One can obtain the original Chen-Lubensky model if $\Dpl=\Dpp=0$ and $\Dpr\neq 0$.
%
%
With $\Cpl=B_{\para}/(2\Psi_0^2 q_0^2)$, $\Cpr=B_{\perp}/(2\Psi_0^2 q_0^2)$ and $\Dpl=\Dpp=\Dpr=0$,
the Chen-Lubensky model is reduced to the Landau-de Gennes model \Eq{L-deG}.
\begin{table*}[htb]
\begin{tabular*}{16cm}{@{\extracolsep{\fill}}l  l}
\hline
\hline
model type & parameter list\\
\hline
original Chen-Lubensky model~\cite{ChenLubensky} & $\Dpl=\Dpp=0$, $\Dpr\neq 0$\\
model of~\cite{Luk'yanchuk} & $\Dpl=\Dpp=\Dpr\neq 0$\\
model of~\cite{Kundagrami} (in the vicinity of $q_{\para}=q_0$) & $\Cpl=4\Dpl q_0^2 \neq 0$, $\Dpp, \Dpr\neq 0$\\
Landau-de Gennes model \Eq{L-deG} & $\tilde{C}_i=B_{i}/(2\Psi_0^2 q_0^2) \,\,\, (i=\para, \perp)$, $\Dpl=\Dpp=\Dpr=0$\\
\hline
\hline
\end{tabular*}
\caption{Relationships between the generalized Chen-Lubensky and other models with the corresponding parameter lists.}
\label{tab:CL}
\end{table*}
\par
The equilibrium director in the Sm-C phase tilts against the layer normal at the angle,
\eq
\alpha=\tan\inv\lrS{\f{1}{q_0}\sqrt{\f{-\Cpr}{2\Dpr}}}.
\label{eq:alpha}
\qe
\par
Next we consider a perturbation of the uniform equilibrium configuration with the generalized Chen-Lubensky model.
The director perturbation is given by
\eq
\delta\bn(\br)=\bn(\br)-\bN,
\label{eq:cndecomp}
\qe
where $\bN$ is the equilibrium director and equals $\lrS{\sin\alpha\cos\psi, \sin\alpha\sin\psi, \cos\alpha}$.
The azimuthal angle $\psi$ is the Goldstone mode in the Sm-C phase.
The layer deformation is expressed only by $u(\br)$
\eq
\Psi(\br)=\Psi_0 \exp\lrL{\f{iq_0}{N_3}\lrS{z-u(\br)}}.
\label{eq:clayer}
\qe
%
\par
The equilibrium condition \Eq{normal}, equivalent to $\bn \para \delta \t{F}/\delta \bn$, reads
\seq
\fder{\t{F}}{n_i}-\f{N_i+\delta n_i}{N_3+\delta n_3}\fder{\t{F}}{n_3} &=& 0\,\,\,(i=1, 2).
\label{eq:normal2}
\sqe
The perturbation expansion requires
\eq
\gamma_i \fder{\t{F}}{n_3}^{(0)}=\fder{\t{F}}{n_i}^{(1)}-\gamma_i\fder{\t{F}}{n_3}^{(1)} - \f{\delta n_i}{N_3}\fder{\t{F}}{n_3}^{(0)}  =0 \,\,\,\, (i=1, 2),
\label{eq:ESAC}
\qe
where $\gamma_i=N_i/N_3\,(i=1,2)$, and $O^{(n)}$ means the $n$th order perturbative part of $O$ in terms of $\delta\bn$ and $u$.
\EEEq{ESAC} has two possible solutions.
We obtain the Sm-A phase ($\bN=\be_3$) if $\lrS{ \delta \t{F}/\delta n_3 }^{(0)} \neq 0$,
otherwise the Sm-C phase with the director tilt angle $\alpha$ \Eq{alpha}.
%
\subsection{Effective Sm-A free energy}
\par
The Sm-A case is quite simple: $\gamma_1=\gamma_2=0$ and $N_3=1$.
%
%
Following the same procedure as in Sec. II, the effective free energy is calculated as
\eq
\delta n_i(\bq) &=& -\f{2i(\t{C}_{\perp} + \t{D}_{\para}q_{\para}^2 + \t{D}_{\perp}q_{\perp}^2)}
{2\t{C}_{\perp} + \Dpd q_{\perp}^2 + q^2} q_i u(\bq) \,\,\,\, (i=1, 2),
\,\,\,\,\,\,\,\,\,\,\, \delta n_3(\br) = 0,
\label{eq:CLAn12}
\\
%
%
\tF&=& 2 \int_{\bq}
\biggl[ \t{C}_{\para}q_{\para}^2 + \t{D}_{\para}q_{\para}^4
+ 2\t{D}_{\para\perp}q_{\para}^2 q_{\perp} + \t{D}_{\perp}q_{\perp}^4 \biggr.\nn\\
&&\biggl.
+ \f{\t{C}_{\perp}q_{\perp}^2\lrS{ \Dpd q_{\perp}^2 + q^2}
-2q_{\perp}^2\lrS{\t{D}_{\para}q_{\para}^2 + \t{D}_{\perp}q_{\perp}^2}\lrS{2\t{C}_{\perp} + \t{D}_{\para}q_{\para}^2 + \t{D}_{\perp}q_{\perp}^2}}
{2\t{C}_{\perp} + \Dpd q_{\perp}^2 + q^2} \biggr] |u(\bq)|^2,
\label{eq:CLAF}\nn\\
\qe
where $\int_{\bq}[\cdots] \equiv \int d\bq[\cdots]/(2\pi)^3$
and $\Dpd \equiv 2\Dpr+\tk1$.
%
The twist and bend contributions of Frank energy do not appear in the effective free energy
because $\delta \bn_{\perp} \para \bq_{\perp}$.
%
The layer compression term has no director contribution,
because the layer width strain energy should be the quadratic form $\D_z u+\delta n_z$,
and the higher order term $\delta n_z$ is neglected. 
\par
Next we compare the previous Landau-de Gennes and the present Chen-Lubensky model briefly,
to understand the role of the fourth order gradient terms.
By setting $\t{D}_{\para}=\t{D}_{\para\perp}=\t{\kappa}_1=0$, \EEq{CLAF} is reduced to the simpler form
\eq
\tF&=& 2\int_{\bq} \lrL{
\t{C}_{\para}q_{\para}^2
+  \t{\kappa}(\bq) q_{\perp}^4
+  \f{\t{C}_{\perp} + \Dpr\qp^2} {2\t{C}_{\perp}+2\t{D}_{\perp}q_{\perp}^2+q^2} q_{\para}^2 q_{\perp}^2 } |u(\bq)|^2,\nn\\
&&\lrS{ \t{\kappa}(\bq) = \f{\t{C}_{\perp} + 2\t{D}_{\perp}\t{C} + \t{D}_{\perp} q_{\perp}^2} {2\t{C}_{\perp}+2\t{D}_{\perp}q_{\perp}^2+q^2} }.
\label{eq:CLAF2}
\qe
\begin{figure}[htbp]
 \includegraphics[width=60mm]{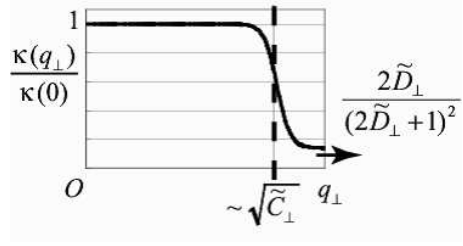}
 \caption{Plot of the effective layer undulation modulus $\tilde{\kappa}(q_3=0, \,\qp)$ in \Eq{CLAF2}.
	The Lorentzian curve with the offset $(\sim \Dpr)$ decays with the characteristic wave number $\sim \sqrt{\Cpr}$.
	The fourth-order gradient coefficients are typically quite small than the second order coefficients.}
 \label{fig:DLo}
\end{figure}
This is identical with the effective Landau-de Gennes free energy \Eq{effL-deG}
if $\t{D}_{\perp}=0$.
The profile of $\t{\kappa}(\bq)$ is depicted in \Fig{DLo}.
The layer bending modulus only from the density contribution
can be considered as $\tilde\kappa(\qp \rightarrow\infty)=2\Dpr/(2\Dpr+1)$
because the splay vanishes at $\qp\rightarrow \infty$.
%
%
\subsection{Effective free energy for Sm-C phase}
\par
Next we calculate the effective free energy for the Sm-C phase.
Using
$\delta \t{F}^{(0)}/\delta n_3=-4\lrM{\t{C}_{\perp}+2\t{D}_{\perp}q_0^2\lrS{(1/N_3^2)-1}}/N_3=0$,
the perturbed equation of state is given in the matrix form
\seq
\bM \delta \bn_{\perp}(\bq) &=& \bv(\bq),
\label{eq:ESC}
\\
\nn\\
\bM &=&
\left(
\begin{array}{cc}
\t{D}^{'}\t{q}_2^2 + \lrS{1-N_2^2}q^2 + 2N_1^2 A + \t{\kappa}_2 \qpr{2}^2
   &  -\t{D}^{'}\t{q}_1\t{q}_2 + N_1 N_2 \lrS{q^2 + 2A} - \t{\kappa}_2\qpr{1}\qpr{2} \\
-\t{D}^{'}\t{q}_1\t{q}_2 + N_1 N_2 \lrS{q^2 + 2A} - \t{\kappa}_2\qpr{1}\qpr{2}
   &  \t{D}^{'}\t{q}_1^2 + \lrS{1-N_1^2}q^2 + 2N_2^2 A + \t{\kappa}_2 \qpr{1}^2
\end{array}
\right),
\nn\\
\\
\bv(\bq) &=& 2i
\left(
\begin{array}{c}
E\gamma_1 + \t{d} \t{q}_2 \\
E\gamma_2 - \t{d} \t{q}_1 \\
\end{array}
\right)
u(\bq),
\sqe
where we use the first order terms of the normalization condition $\delta n_3=-\gamma_1 \delta n_1-\gamma_2 \delta n_2$.
Here we introduced the abbreviations
\seq
A &\equiv& \f{1}{N_3^2}\lrS{\t{C}_{\para} + 4\t{D}_{\perp}q_0^2 + \f{2\t{D}_{\para\perp}q_0^2 \Np^2}{N_3^2}}
\label{eq:A}
,\\
E &\equiv& \f{4\t{D}_{\perp}q_0^2 \t{q}_3}{N_3} - \lrS{\t{C}_{\para} + \f{2\t{D}_{\para\perp}q_0^2 \Np^2}{N_3^2}}\bN\cdot\bq,\\
\t{\bq} &\equiv& \bq\times\bN,\\
\t{\bq}_{\perp} &\equiv& \bq-\bN\lrS{\bN\cdot\bq}=\bN\times\t{\bq},\\
\t{d} &\equiv&  \t{D}_{\para\perp} \lrS{\bN\cdot\bq}^2 + \t{D}_{\perp} \lrS{\bN\times\bq}^2.
\sqe
The smectic free energy is invariant under a uniform layer displacement,
and the solution of the equation of state \Eq{ESC} for $q=0$ is $\bN\cdot\delta \bn_{\perp}=\delta n_3=0$.
This is nothing but a uniform rotational Goldstone mode of the $\bc$-vector.
\par
Inverting the matrix $\bM$, we obtain the expression for the director deformation
\eq
\left(
\begin{array}{c}
\delta n_1(\bq)\\
\delta n_2(\bq)\\
\end{array}
\right)
&=&
\f{2iu(\bq)}{D_C}
\left(
\begin{array}{r}
%
%
-N_3\lrS{E\t{D}^{'}\t{q}_2 + 2\t{d}AN_2}\t{q}_3
	+ EN_3N_1 q^2\\
- E\k2\qpr{1}\qpr{3} - N_3\t{d}\qpr{1}\lrS{q^2+\k2\t{q}^2}\\
N_3\lrS{E\t{D}^{'}\t{q}_1 + 2\t{d}AN_1}\t{q}_3
	+ EN_3N_2 q^2\\
- E\k2\qpr{2}\qpr{3} - N_3\t{d}\qpr{2}\lrS{q^2+\k2\t{q}^2}
\end{array}
\right),
\label{eq:CLCn}
\qe
where the determinant of $\bM$ is given by
\eq
D_C &=& N_3^2\lrL{\lrS{\t{D}^{'}+\t{\kappa}_2}q^2 \t{q}^2 + \t{D}^{'}\t{\kappa}_2 \t{q}^4
+2\t{D}^{'}A\t{q}_3^2 + 2A\t{\kappa}_2\qpr{3}^2 + 2N_{\perp}^2q^2A + q^4}.
\qe
Now in the Sm-C phase, $\delta\bn \nparallel \bqp$
even at the pure undulation ($q_3=0$).
So not only the splay Frank elastic energy but also the twist and bend terms have contributions to the layer bending elastic energy.
The splay term
favors the state with $\bq \perp \delta \bn$ (we call it the $\bq \perp \delta \bn$ state),
and both the twist and the bending terms favor the $\bq \para \delta \bn$ state.
In such a layered smectic phase with the spontaneously symmetry breaking layer normal direction
and with the rotational Goldstone mode of the $\bc$-vector,
the three Frank elastic terms
have anisotropic energy contributions,
together with the anisotropic coupling constants.
The leading order terms including $\Dpl$, $\Dpr$ and $\Dpp$ are
$(\lap_{\para}u)^2$, $(\lap_{\perp}u+\div{\bn})^2$ and $(\lap_{\para}u)(\lap_{\perp}u+\div{\bn})$,
which favor $\bN \perp \bq$, $\bN \para \bq$ and the intermediate state respectively.
\par
Anisotropy due to the $\bc$-vector is apparent in the pure undulation case ($q_3=0$).
Let $\vartheta$ be the angle between $\bc$ and $\bqp$.
The effective free energy is calculated with \Eq{CLCn},
\seq
\tF&=& 2 \int_{\bq_{\perp}} \t{f} = 4 \int_{\bq_{\perp}} q_{\perp}^4|u(\bq)|^2 \lrS{\kappa_u(x) + \kappa_n(\bq, x)},
\label{eq:fx}
\\
\nn\\
\nn\\
\kappa_u(x) &=&
\f{ \t{D}_{\para} N_{\perp}^4x^2 + 2\t{D}_{\para\perp}N_{\perp}^2xy + \t{D}_{\perp}y^2} {N_3^2}, \\
%
\kappa_n(\bq, x) &=&
\f{1}{N_3^2}
\f{ \t{D}_{n1}(x) N_3^2\qp^2\lambda_A^2 + \t{D}_{n2}(x) N_{\perp}^2 }{ \t{D}_{n3}(x) N_3^2\qp^2\lambda_A^2 + \t{D}_{n4}(x) N_{\perp}^2 },\\
\sqe
\seq
\t{D}_{n1}(x) &=& -2\lrS{\t{D}_{\para\perp}x+\t{D}_{\perp}y}^2 y \lrS{1+\t{\kappa}_2y}, \\
\t{D}_{n2}(x) &=& xN_3^2\lrM{ 1 + \t{D}\uprime y
 -4\lrS{\t{D}_{\para\perp}x+\t{D}_{\perp}y}} \lrS{1+\t{\kappa}_2y}, \nn\\
&& - 4\lrS{\t{D}_{\para\perp}x+\t{D}_{\perp}y}^2 \lrS{1-x}, \\
\t{D}_{n3}(x) &=& \lrS{1+\t{D}\uprime y} \lrS{1+\t{\kappa}_2 y}, \\
\t{D}_{n4}(x) &=& 2 \lrS{ \t{D}\uprime\lrS{1-x} + \t{\kappa}_2N_3^2 x + 1},
%
%
%
%
\sqe
where $x \equiv \cos^2 \vartheta$ and $y(x)=1-N_{\perp}^2x$.
The effective layer bending elasticity contains the two components:
$\kappa_u(x)$ is the density contribution without the director deformation,
and $\kappa_n(\qp, x)$ comes both from the density and director elasticity.
The director part has a new characteristic length-scale $\lambda_A \equiv 1/\sqrt{A}$,
determined by the ratio of the Frank elasticity and the combination of the second and fourth order density gradient terms \Eq{A}.
Taking the Landau-de Gennes limit $\t{D}_i  \rightarrow 0\,\,(i=\perp, \para\perp, \para)$ and $\alpha \rightarrow 0$,
$\lambda_A$ is reduced to the parallel penetration length $\lambda_{\para}=\sqrt{\Cpl}$.
It is completely different from the Sm-A case where the characteristic length is $\lambda_{\perp}=\sqrt{\Cpr}$ \Eq{CLAF2}.
\par
We will compare this result with the previous works in Sec. III E.
\subsection{State diagram for Sm-C phase}
In the following, we examine the minimizer $\vartheta=\theta$ for $\t{F}(\vartheta)$ \Eq{fx}.
The angles $\vartheta$ and $\theta$ are spatially uniform by assumption.
%
%
\par
We first consider the one Frank constant case: $K_1=K_2=K_3$.
In this case, the calculation and evaluation of the free energy become quite simple with no further approximation.
One can show that the free energy density $\t{f}(x)$ is a convex function throughout $0\leq x \leq 1$,
because a stability of the smectic layer requires the inequality $\det(\t{D}) \equiv \Dpl\Dpr-\Dpp^2>0$ (see Appendix B).
Thus the behavior of the free energy minimum is determined only by $\t{f}^{'} (0)$, $\t{f}^{'} (1)$ and $\det(\t{D})$.
%
There are three possible cases for the first derivative $\t{f}^{'} (x)$:\\
(i) $\t{f}^{'} (0)>0$ (so $\t{f}^{'} (1)>0)$\\
\, The minimum of the free energy is at $\theta=90\deg$\\
(ii) $\t{f}^{'} (0)<0$ and $\t{f}^{'} (1)>0$\\
\, The favored angle $\theta$ exists between $0\deg$ and $90\deg$.\\
(iii) $\t{f}^{'} (0)<0$ and $\t{f}^{'} (1)<0$.\\
\, The free energy minimum is at $\theta=0\deg$, and the director tends to be aligned to the layer bending direction.
\begin{figure}[htbp]
 \leavevmode
 \begin{center}
  \begin{tabular}{ lcl }
	(a) $\alpha=10\deg, \qp=2\times10^6$cm$\inv$
	&& (b) $\alpha=10\deg, \qp=2\times10^7$cm$\inv$
\\
	\includegraphics[width=40mm]{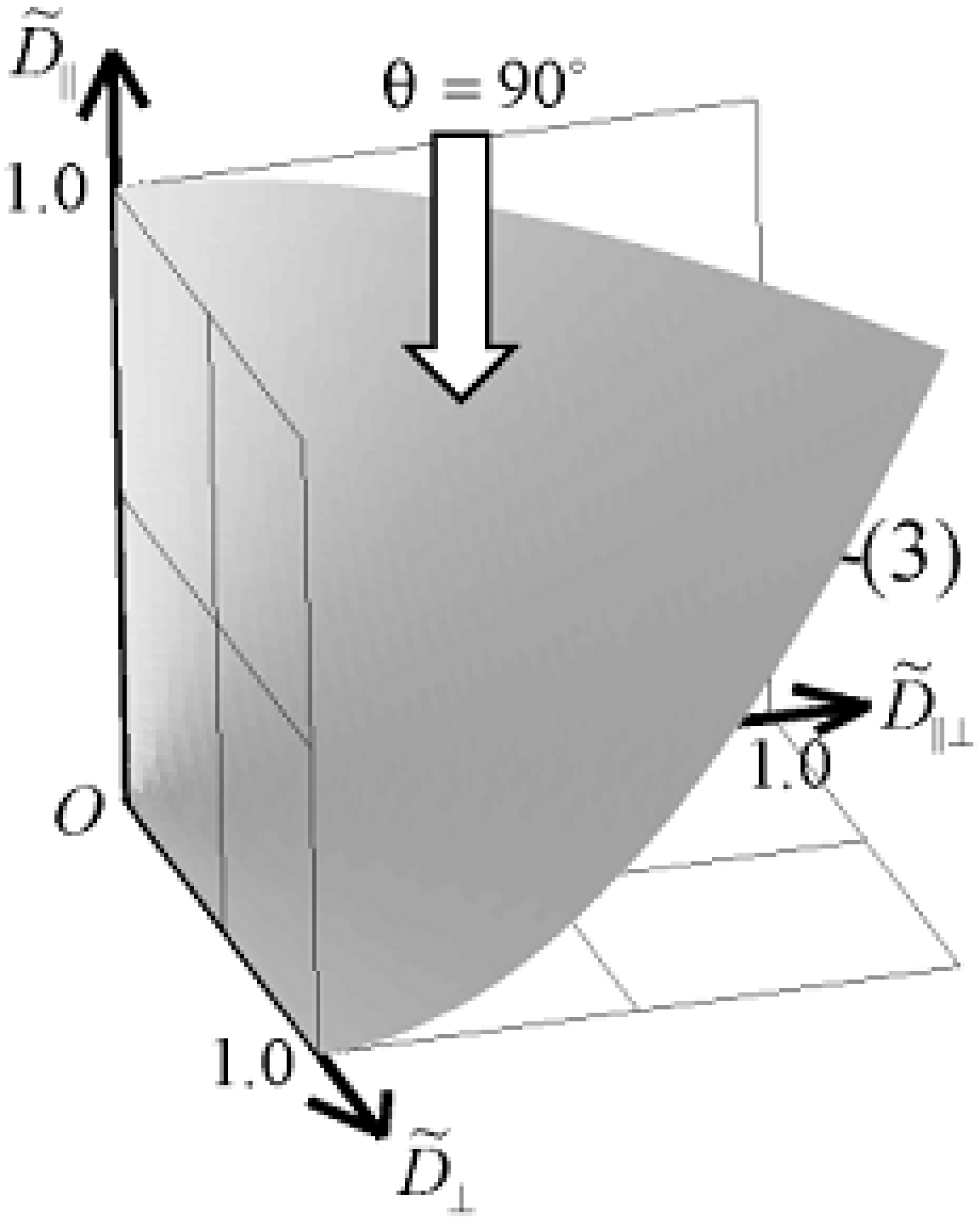}
	&&
	\includegraphics[width=40mm]{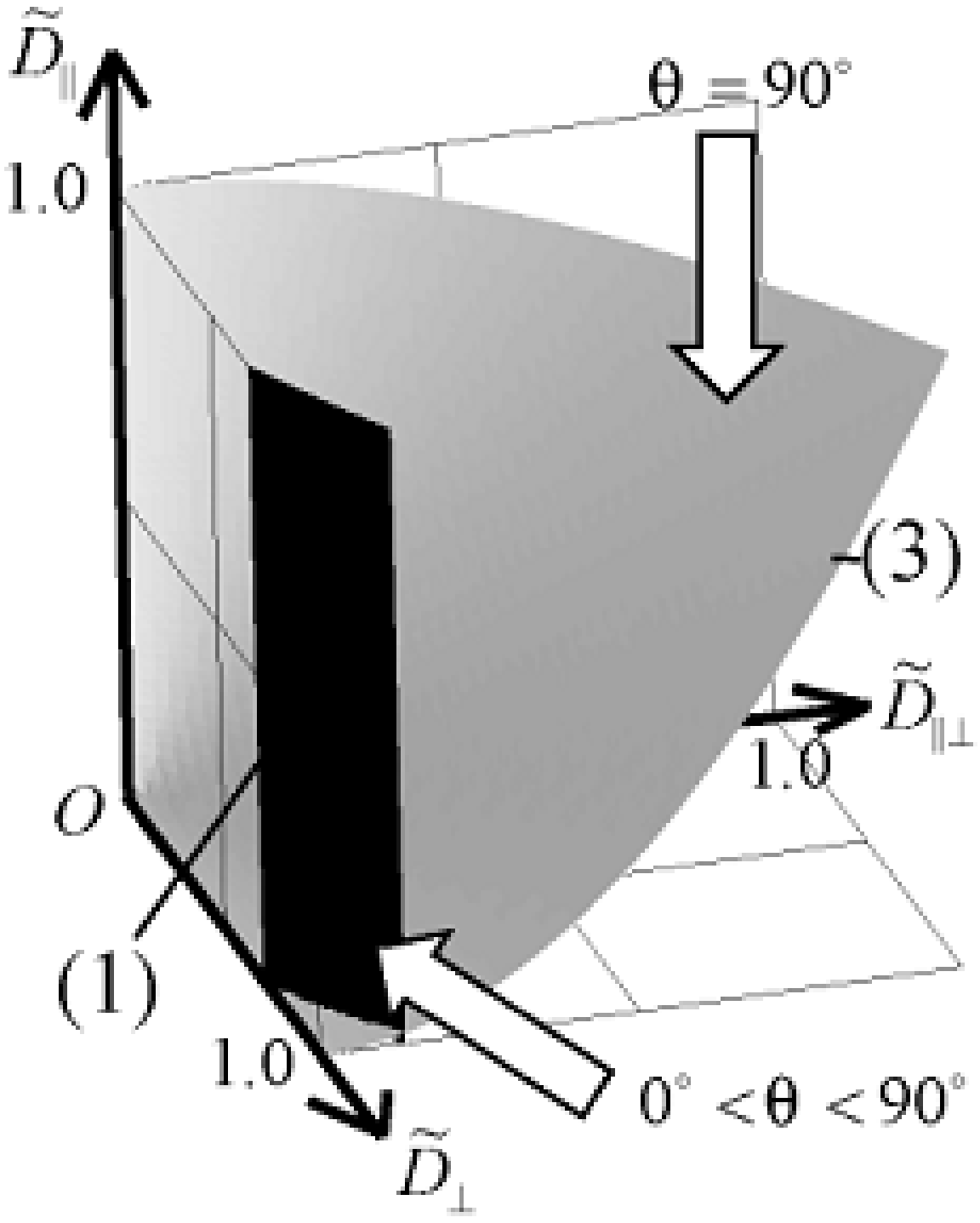}
\\
	(c) $\alpha=30\deg, \qp=2\times10^6$cm$\inv$
	&& (d) $\alpha=50\deg, \qp=2\times10^6$cm$\inv$
\\
	\includegraphics[width=40mm]{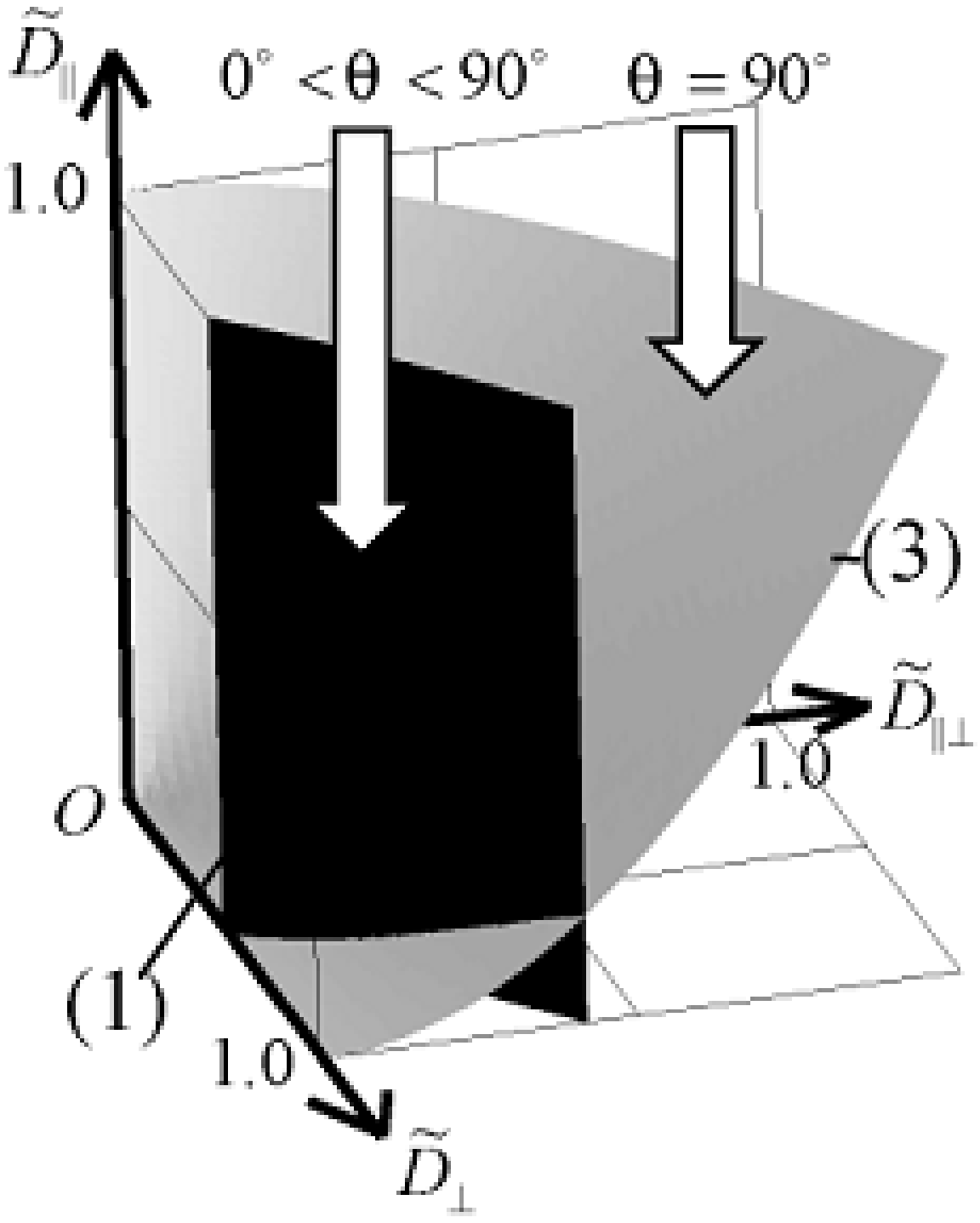}
	&&
	\includegraphics[width=40mm]{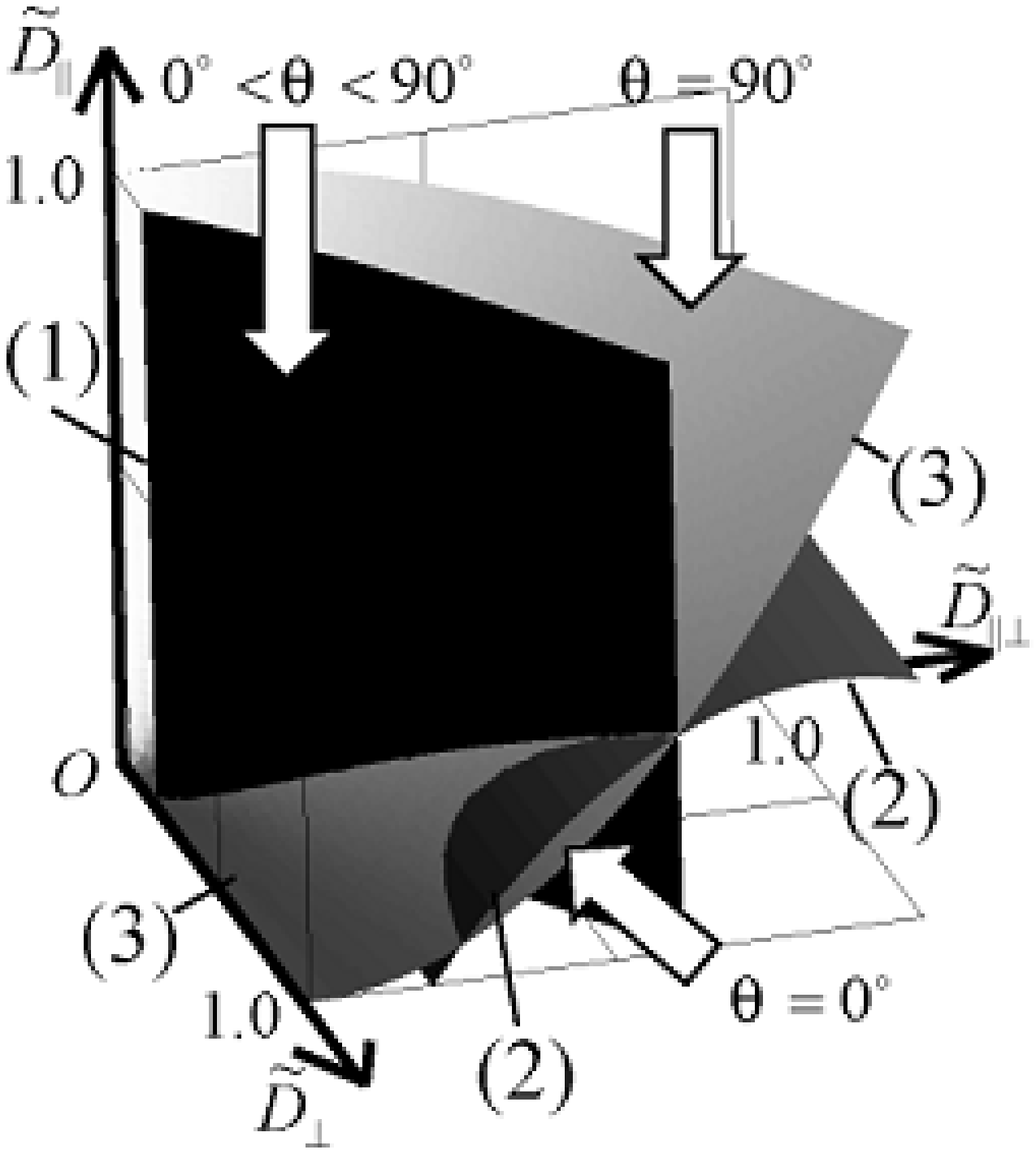}

  \end{tabular}
  \caption{State diagrams with
$\Dpr$, $\Dpp$ and $\Dpl$
in the one Frank constant approximation.
	We use
$q_0=2\times10^7$cm$^{-2}$ and
	$\Cpl=4\times10^{12}$cm$\inv \lrS{ \sim \lambda_{\para}^{-2}}$ as a set of typical values~\cite{Renn, NavaillesTGBC}.
	The planes $\t{f}^{'}(0)=0$, $\t{f}^{'}(1)=0$ and $\det(\t{D})=\Dpr\Dpl-\Dpp^2=0$
	are denoted by the plane (1), (2) and (3) respectively.
	}
	%
  \label{fig:1FSDA}
 \end{center}
\end{figure}
%


\par
The fixed parameters are set to be typical values and used in followings if not specified:
$\qp=2\times 10^7$cm$\inv$
and
$\Cpl=\lambda^{-2}_{\para}=4\times 10^{12}$cm$^{-2}$~\cite{Renn, NavaillesTGBC}.
We limit the possible parameter range of $\lrM{ \t{D}_i \,|\, i=\para, \para\perp, \perp}$ to $0 < \t{D}_i <1$,
because of the stability of the free energy, the relation $\Dpr \sim d^2/(\lambda_{\perp}\Np)^2$ holds,
and the feasibility of the Landau-de Gennes model $(\t{D}_i=0)$ in the Sm-A phase.
In fact, one of the scattering experiments shows that $\Dpp$, $\Dpl \simeq 0$ and $\Dpr(\sim 10^{-4}) \ll 1$~\cite{Martinez},
where only the trivial $\theta=90\deg$ state is expected to be observed as the previous work implies~\cite{Hatwalne, Johnson}.
However, depending on the material, one might be able to obtain a wide range of the parameter sets within $0 < \t{D}_i \alt 1$.
The three-dimensional state diagrams for $\Dpl$, $\Dpp$ and $\Dpr$
are given in \Fig{1FSDA}.
State boundaries are determined by the conditions (1) $\t{f}^{'} (0)=0$, (2) $\t{f}^{'} (1)=0$,
and (3) $\det(\t{D})=0$.
%
\par
We again note that the smectic layer elasticity consists only of
the density and the Frank elastic energies and the coupling energy between the layer normal and the director.
All the following results will be explained in terms of combinations of the three elastic contributions.
\par
The $\theta=90\deg$ state dominates for small $\alpha$ and $\qp<\lambda_A\inv (\sim 10^7$cm$\inv)$ (\Fig{1FSDA}(a)).
When $\qp \agt \lambda_A\inv$, on the other hand, the Frank elasticity less contributes to $\tilde{\kappa}(\bq)$,
and anisotropic density terms dominate (\Fig{1FSDA}(b)),
being enhanced for larger $\alpha$ (\Fig{1FSDA}(c)).
Thus various angle $\theta$ can appear.
In experiment, the largest tilt angle $\alpha$ that has been reported to our knowledge is $37\deg$~\cite{Meiboom}.
So the possible stable $\theta$ state might be limited to $0\deg < \theta \le 90\deg$, as we see from the diagrams for $\alpha \le 40\deg$.
Nevertheless, if the $\alpha=50\deg$ molecular tilt is achieved,
there exists the stable $\theta=0\deg$ state (\Fig{1FSDA}(d)).
%
\par
In each state diagram,
the behavior of $\theta$ is quite sensitive to $\Dpr$
compared with $\Dpp$ and $\Dpl$.
This is because the elasticity of the vector $\t{\bc}$ $\lrS{\simeq -\bc}$
is dominated by the elastic terms including $\Dpr$, as we discussed in Sec.II A.
The ratio between $\Cpr$ and $\Dpr$ is uniquely determined by the tilt angle $\alpha$ \Eq{alpha},
so the value of $\Dpr$ controls the preference of $\bqp \para \bN$.
Thus the stable state changes from $\theta=90\deg$ to $0\deg<\theta<90\deg$ and $\theta=0\deg$
as $\Dpr$ grows.
On the other hand
in \Fig{1FSDA}(d),
the $\theta=0\deg$ state transits to the $0\deg<\theta<90\deg$ state with the increment of $\Dpl$.
\begin{figure}[htbp]
 \leavevmode
 \begin{center}

  \begin{tabular}{lclclcl}
	&& (a) $\alpha=40\deg$
	&& (b) $\alpha=50\deg$
	&& (c) $\alpha=50\deg$, $\Dpr=0.2$\\
	\includegraphics[width=13mm, height=26mm]{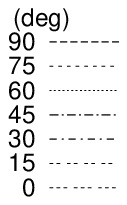}
	&&\includegraphics[width=40mm]{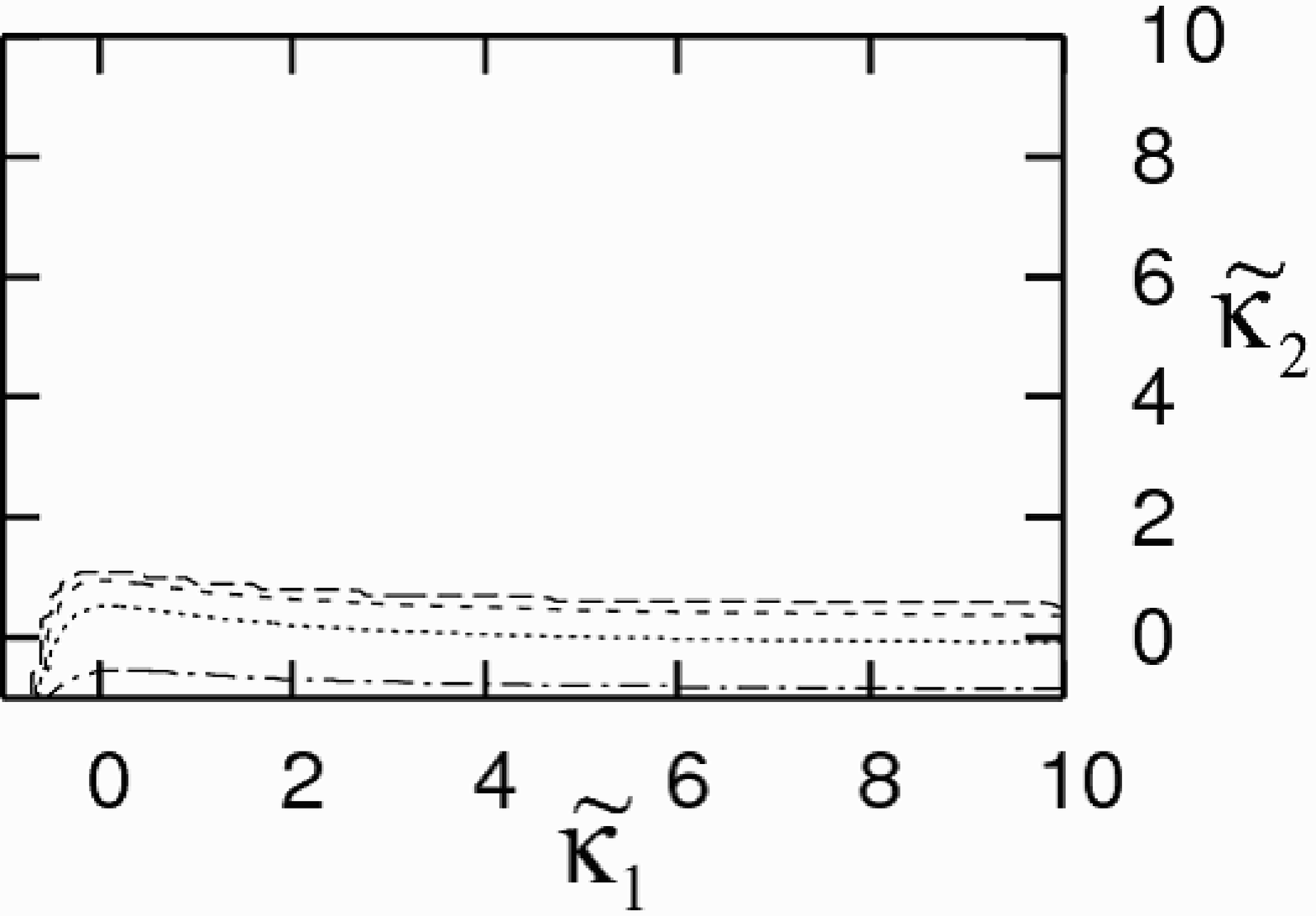}
	&&\includegraphics[width=40mm]{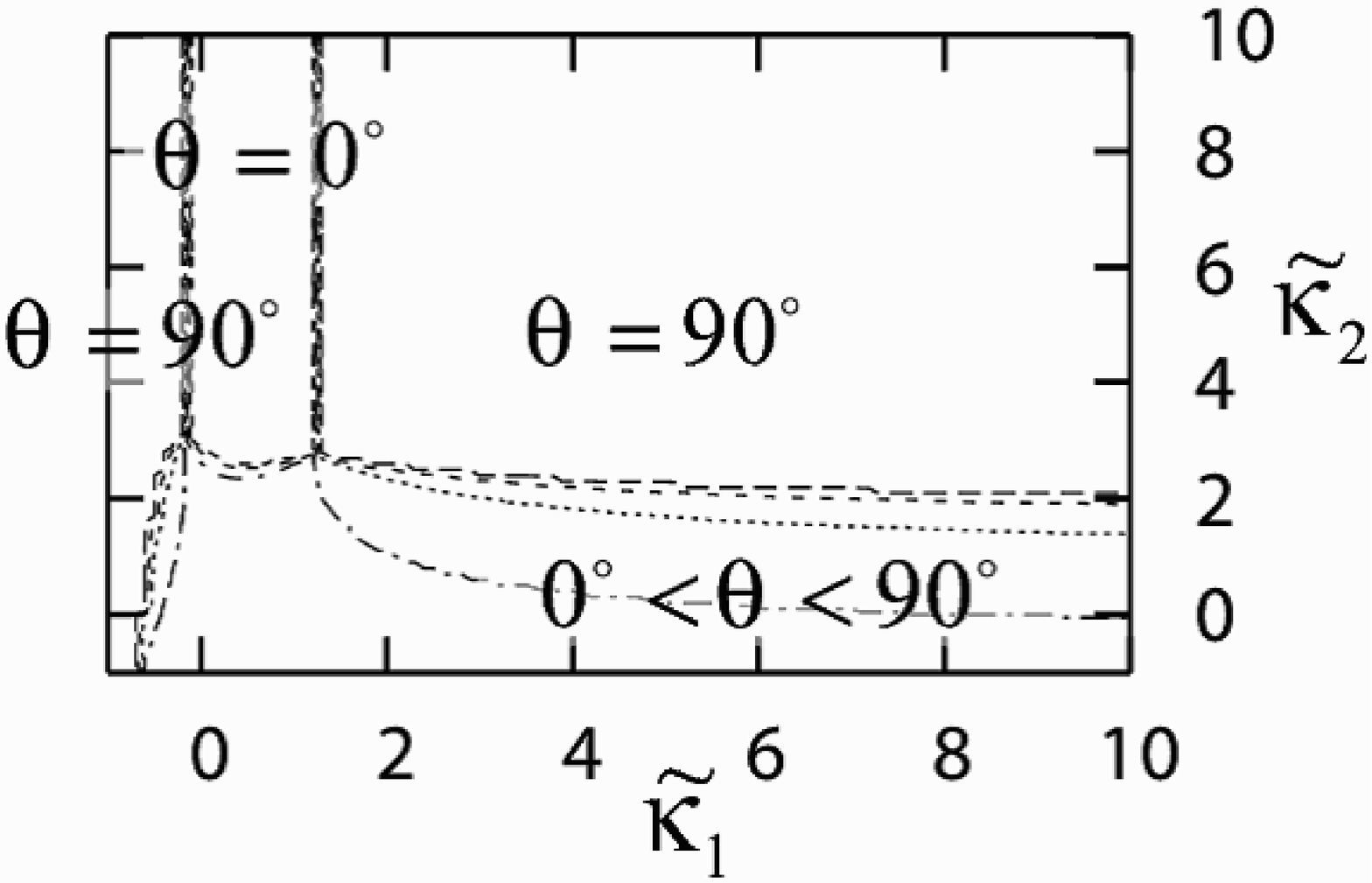}
	&&\includegraphics[width=43mm]{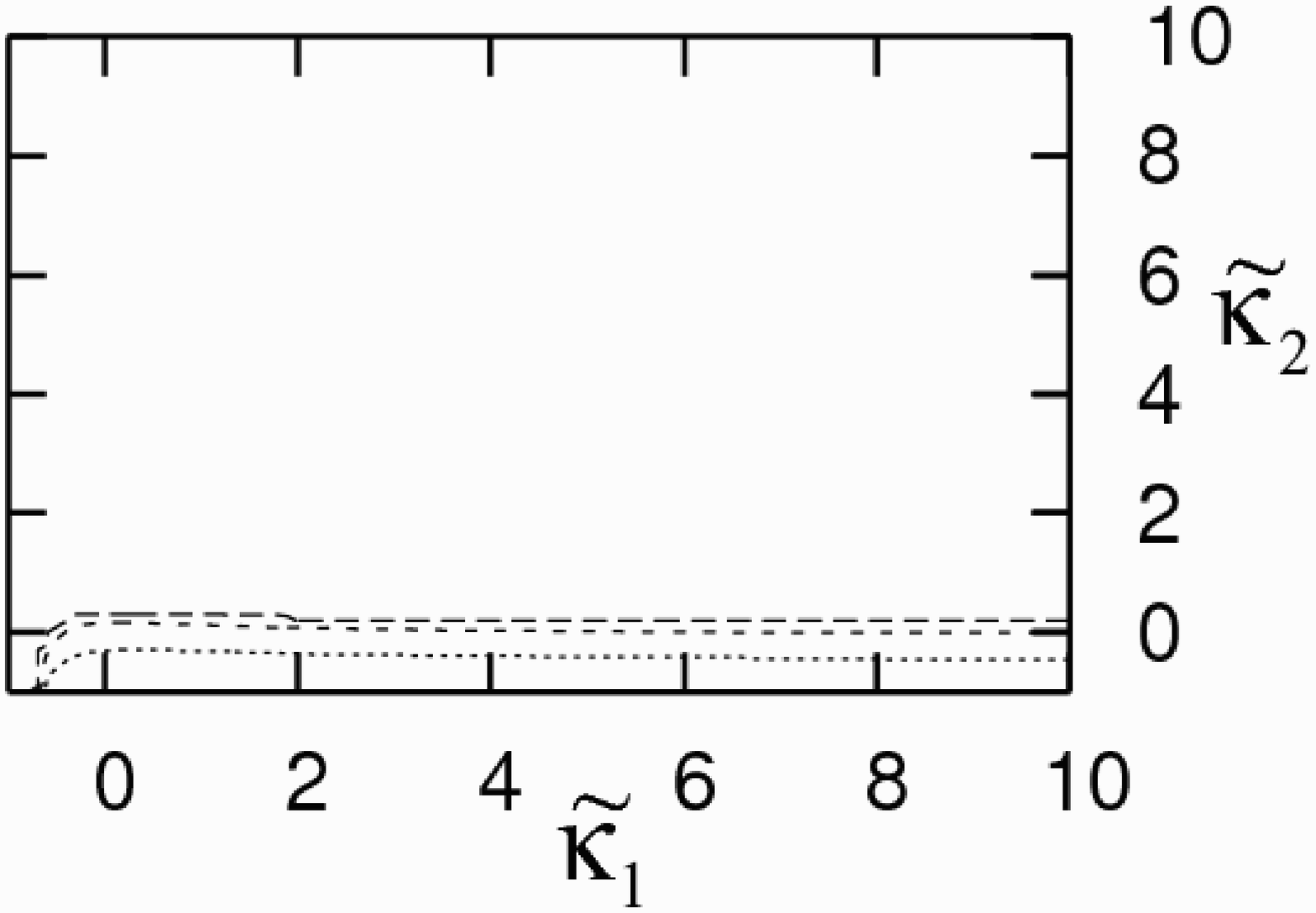}\\
  \end{tabular}
  \caption{State diagrams in terms of
$\tk1$ and $\k2$ with different tilt angles.
	(a) The $0\deg < \theta < 90\deg$ domain gradually disappears as $\alpha \rightarrow 0$.
	For $\alpha=40\deg$, the $0\deg < \theta < 90\deg$ domain grows around $\tk1 \simeq 0$.
	(b) For $\alpha=50\deg$, a stable $\theta=0\deg$ domain emerges at $\tk1 \simeq 0$ and higher $\k2$.
	The preferred angle changes discontinuously between the $\theta=0\deg$ and $\theta=90\deg$ states. The lower $\Dpr$ diagram is plotted in (c).}
  \label{fig:aFka}
 \end{center}
\end{figure}
\begin{figure}[htbp]
 \leavevmode
 \begin{center}
%
%
  \begin{tabular}{lclcl}
	(a) $\alpha=20\deg$ (solid line)
	&& (b) $\alpha=50\deg$, $\tk1=0$, $\k2=6$\\
	\hspace{5.3mm}$\alpha=40\deg$ (dashed line)
	&& ($K_1/K_3=1$, $K_2/K_3=7$)\\
	\includegraphics[width=40mm]{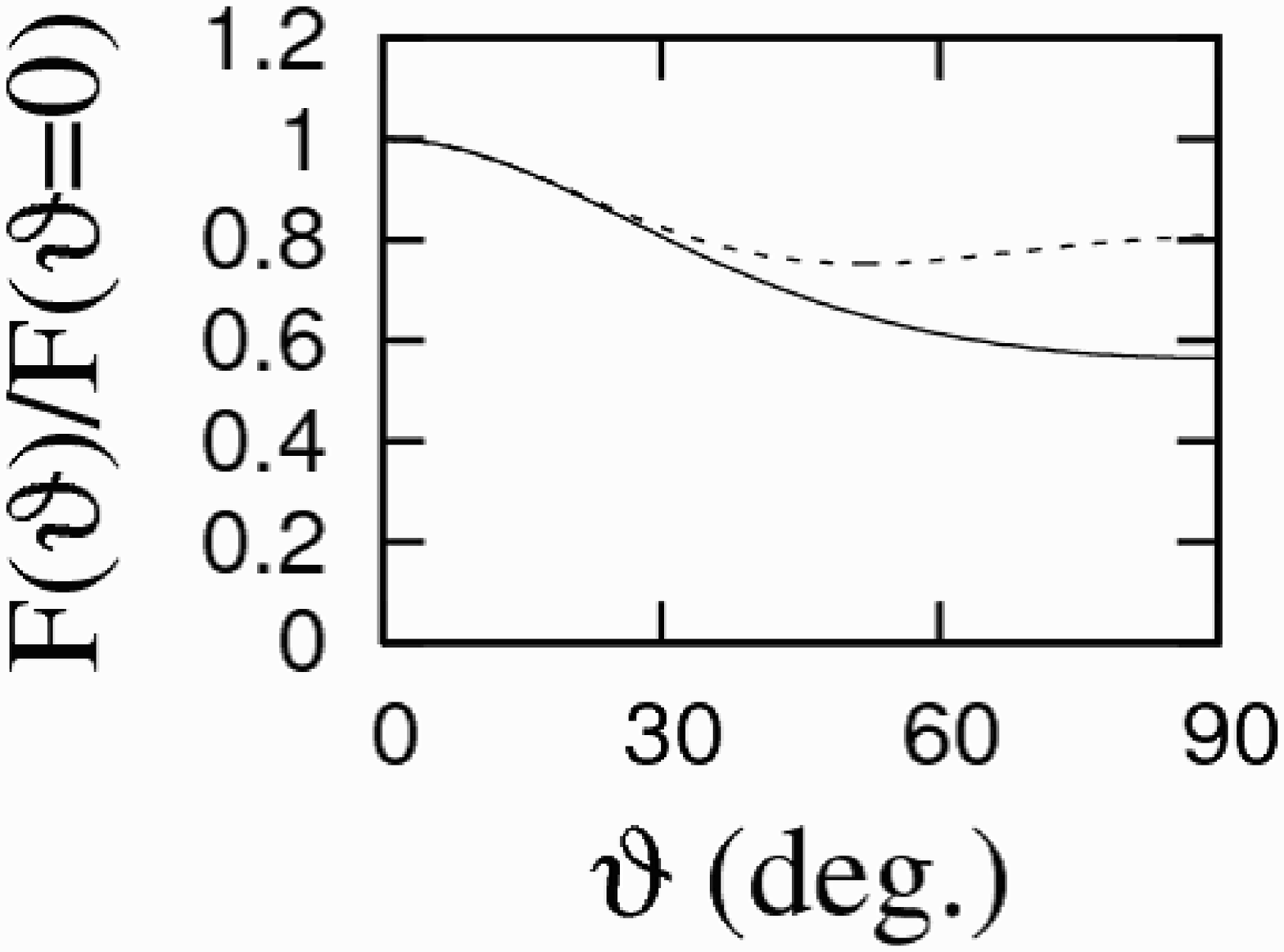}
	&&\includegraphics[width=40mm]{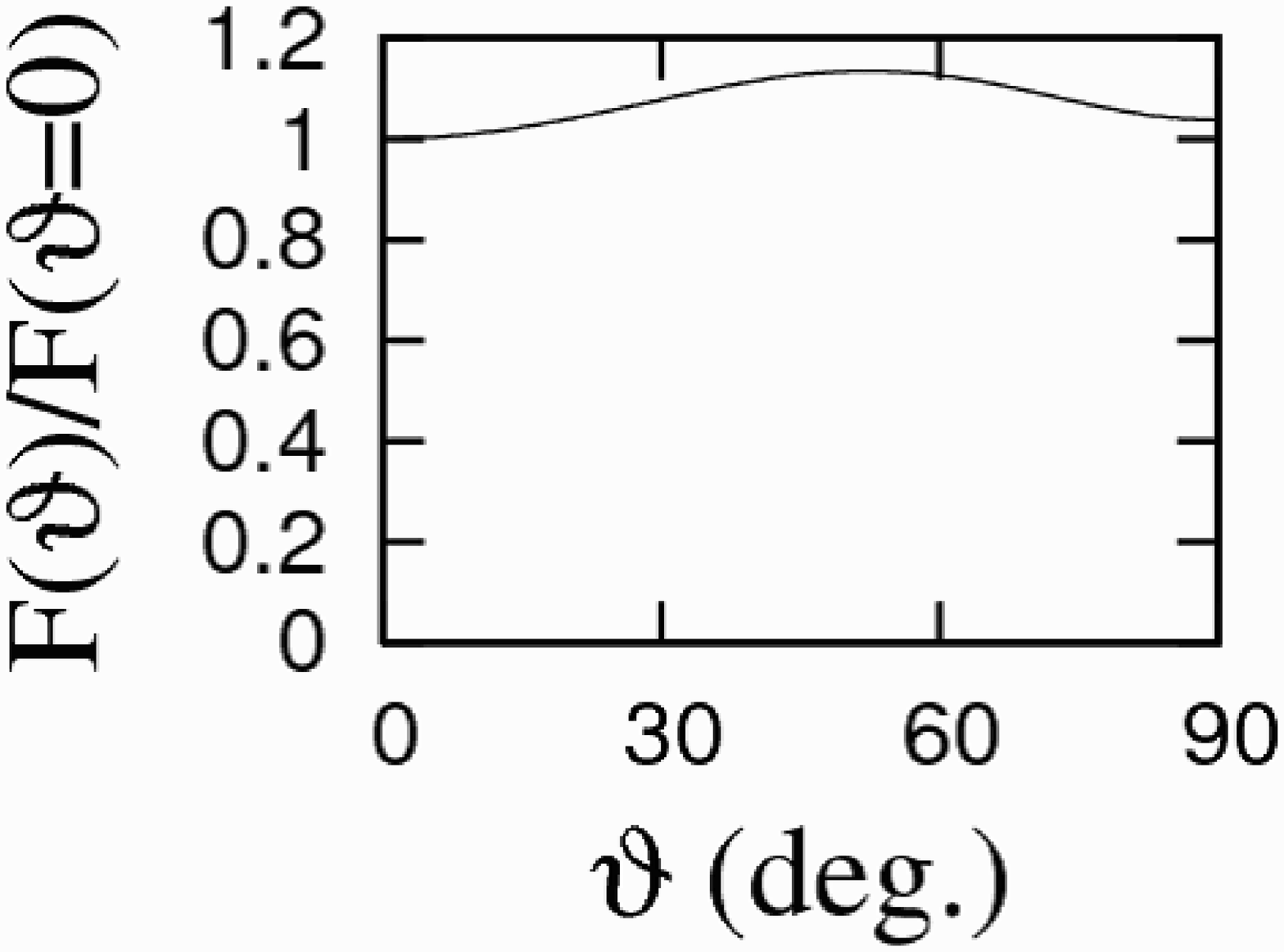}
  \end{tabular}
  \caption{Plot of the free energy $f(\vartheta)$
			as a function of the angle between the $\bc$-vector and $\bqp$.
			(a) The tilt angle dependence of the free energy $f(\vartheta)$.
			(b) Discontinuous transition is explained with the two minima at $\theta=0\deg$ and $ 90\deg$
			as it should be.
	}
  \label{fig:Faa}
 \end{center}
\end{figure}
\par
The state diagram for the reduced Frank elastic moduli $\tk1$ and $\k2$ is also calculated based on \Eq{fx}.
We use $\Dpr=0.5$, $\Dpp=0.01$ and $\Dpl=0.1$.
An experimental result tells that $\Dpp$ is smaller than $\Dpr$, $\Dpl$~\cite{Martinez}.
$\Dpr = 0.5$ is about on the state boundary between $\theta=90\deg$ or less,
so that we would obtain the sensitive state behavior.
This parameter set will be used in the following if not specified.
%
\par
Effect of the Frank elastic constants is quite sensitive to $\alpha$.
%
\begin{figure}[htbp]
 \leavevmode
 \begin{center}
  \begin{tabular}{lcl}
	(a) $\vartheta=0\deg$
	&& (b) $\vartheta=90\deg$\\
	\includegraphics[width=41mm]{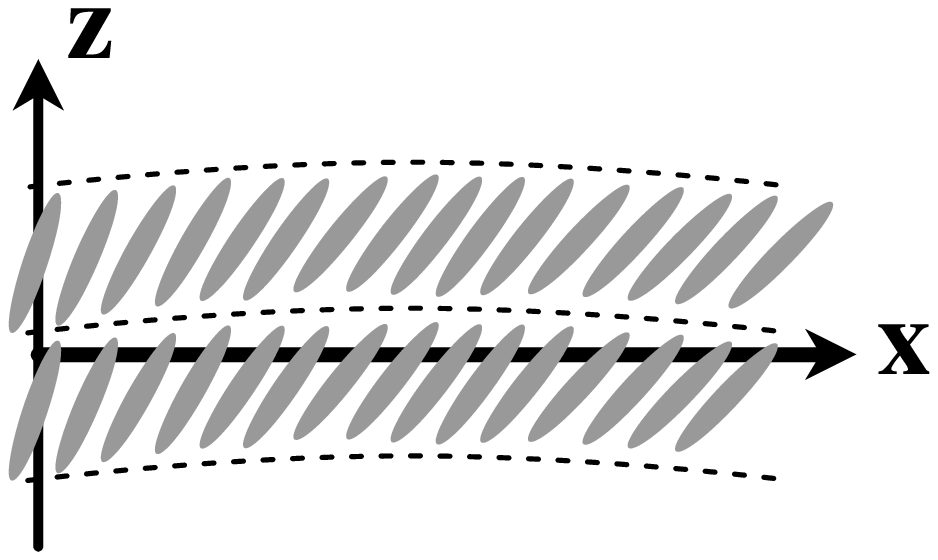}
	&&\includegraphics[width=41mm]{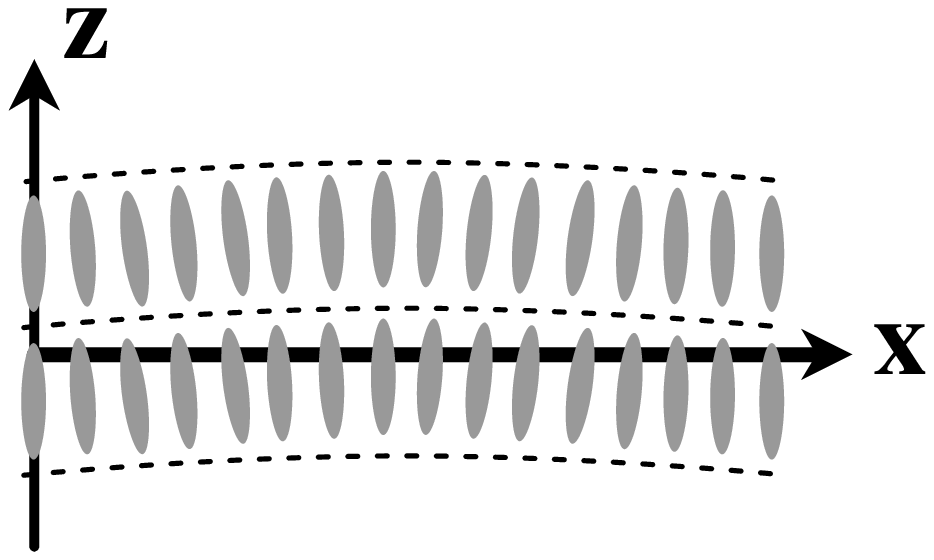}
  \end{tabular}
  \caption{Layer undulation and the director configuration assuming the nonzero tilt angle $\alpha$.
The splay Frank elastic energy costs at $\vartheta=0\deg$ more than at $\vartheta=90\deg$,
because of the director component projected to the $zx$-plane.}
  \label{fig:splay}
 \end{center}
\end{figure}
%
For $\alpha\alt 40\deg$ (\Fig{aFka}(a)), the $\theta=90\deg$ state dominates except for low $\k2$,
while $\theta<90\deg$ domain shrinks for smaller and larger $\tk1$.
The $\theta=90\deg$ state is achieved by a combination of the locking term
(which is the major contribution for small $\alpha$),
and the twist Frank term $|\bN\cdot\lrS{\bqp\times\delta\bn}|^2$
favoring $\bqp\para\bN$ or $\bqp\para\delta\bn$.
The $\theta<90\deg$ domain is suppressed by the two mechanisms.
One for larger $\tk1$ is a combination of the locking term and the splay Frank term (see \Fig{splay} and the detailed calculation is found in Appendix C).
The other for smaller $\tk1$ is a combination of the twist and bend Frank elastic energies,
which favor $\bqp\para\delta\bN$.
For $\alpha>50\deg$ (\Fig{aFka}(b)), the locking term less contributes and the anisotropic elasticity with $\Dpr$ is enhanced more.
Thus both of the two minima $\bqp\para\bN$ and $\bqp\para\delta\bn$ are stable for higher twist $\k2$.
They are incompatible because of the normalization condition $\bN\cdot\delta\bn=0$,
so the transition between $\theta=0\deg$ and $\theta=90\deg$ is discontinuous.
This anomalous state behavior disappears for weak anisotropy (\Fig{aFka}(c)).
\par
We also plot the free energy $f(\vartheta)$ in \Fig{Faa}.
The normalization is taken with $f(\vartheta=0\deg)$.
The minimizer $\theta$ decreases as $\alpha$ grows,
but the free energy barrier is still small for $\alpha=40\deg$,
being a few percent of the absolute value \Fig{Faa}(a).
So the equilibration to the free energy minimum might be disturbed by the internal thermal noise and an external force.
%
%
However the $0\deg<\theta<90\deg$ state is more stable at higher $\alpha$.
%
%
The discontinuous $\theta$ transition in \Fig{aFka}(c) is certainly due to the coexistence of the two local minima (\Fig{Faa}(b)).
\subsection{Discussion}
\par
We next compare our calculation with the previous works.
Hatwalne and Lubensky~\cite{Hatwalne} derived the elastic free energy for the Sm-C phase
in a more elegant way making use of the symmetry and covariance of the model free energy.
The resulting representation in the case $\bN=(N_1, 0, N_3)$ is
\eq
F &=& \f{1}{2} \int_{\bq} \lrS{ B q_3^2 + K_{11} q_1^4 + 2K_{12}q_1^2 q_2^2 + K_{22} q_2^4} |u(\bq)|^2
\label{eq:FHL}
\qe
We can obtain the layer compression coefficient from \Eq{CL}, \Eq{CLCn}, and the normalization $\bN\cdot\delta\bn=0$,
by setting $q_1=q_2=0$ and $q_3 \rightarrow 0$,
\eq
B &=& \f{8 K_3\Dpr q_0^2}{N_3^6 A} \lrM{ AN_3^4\lrS{2-N_3^2} - 4\Dpr q_0^2}.
\qe
We note that the layer compression elasticity stems from the director contribution $\tk1$ as well as the density contribution
in the Sm-C phase,
because a change of the layer width is associated with the change of the tilt angle.
%
\par
As for the layer bending elasticity,
the effective Chen-Lubensky model \Eq{fx} cannot be written in the Hatwalne-Lubensky $q$-dyadic form $\Eq{FHL}$,
because in general the anisotropic layer bending elasticity should be expressed as
\eq
F &=& \f{1}{2} \int_{\bq} \lrS{ Bq_3^2 + \qp^4 K(\vartheta, \bqp)}|u(\bq)|^2.
\qe
Still we can give some representations for $K_{11}$ and $K_{22}$ in the simple form,
\seq
K_{11} &=& K(\vartheta=0\deg) = \f{K_3}{N_3^2} \lrM{ 2\Dpl\Np^4 + N_3^2\lrS{ 1+\tk1 N_3^2 } },\\
K_{22} &=& K(\vartheta=90\deg) = \f{2 K_3 \Dpr \lrS{\tk1+1} }{ N_3^2 \lrS{2\Dpr+\tk1+1} }.
\sqe
$K_{12}$ is calculated with the parabolic approximation,
\eq
K_{12} &\simeq& \f{ K({\vartheta=0\deg}) + K({\vartheta=90\deg}) }{2}
-\f{1}{4} \lrS { \left.\der{K}{x}\right|_{\vartheta=0\deg} - \left.\der{K}{x}\right|_{\vartheta=90\deg} }
\qe
The three elastic constants $K_{11}$, $K_{12}$, and $K_{22}$ contain both the density and the director contributions as pointed out in the previous research~\cite{Hatwalne}.
\par
In this way, we bridge the models of smectic liquid crystal at different coarse-graining levels.
It leads one to a quantitative discussion with the phenomenological macroscopic model~\cite{Hatwalne},
based on the experimental determination of the Chen-Lubensky model~\cite{Martinez}.

\par
We conclude this section comparing our results for Sm-C liquid crystals with other layer forming materials.
Diblock copolymers, chemically connected two different homopolymers,
do not possess the definite molecular orientational degree of freedom especially in the weak segregation regime~\cite{Leibler}.
Thus anisotropic elasticity cannot occur.
On the other hand, in surfactant systems, the linear amphiphilic molecule is basically normal to the surfactant layer~\cite{GompperKlein, ChenJayaprakash}.
This is similar to the Sm-A phase of liquid crystals rather than the Sm-C phase.
Thus amphiphilic system
again would not have an in-plane anisotropic coupling.
However in some cases the surfactant molecules may be aligned at a nonzero tilt angle
against the layer normal~\cite{SafranRobbins, WangGong, KuhnRehage}.
Such systems might have an anisotropic layer elasticity as in the present Sm-C liquid crystal case.
%
\section{Hydrodynamics of Smectic-C layers}
In this section, we consider the transverse wave in smectic liquid crystals which oscillates in the layer normal direction,
and travels perpendicularly to the layer normal.
The wave frequency is much lower than the inverse of the molecular time scale ($\sim 10^8$Hz),
so that the permeation effect which changes the layer width is negligible~\cite{deGennes}.
Also we eliminate some rapidly relaxed degrees of freedom,
namely the density $\rho$, the local energy $\epsilon$,
the velocity along the layer in-plane directions, the pressure $P$, and the $\bc$-vector in the time scale $\sim 10^{-4}$s.
\par
\begin{figure}[htbp]
 \leavevmode
 \begin{center}
  \begin{tabular}{lcl}
	(a)
	&& (b)\\
	\includegraphics[width=45mm]{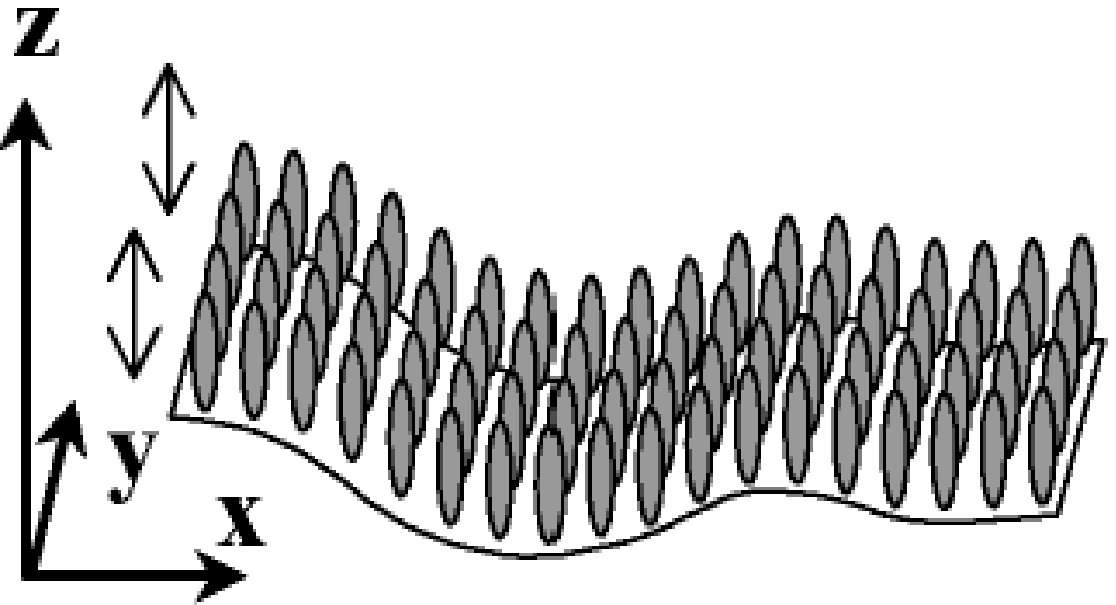}
	&&\includegraphics[width=45mm]{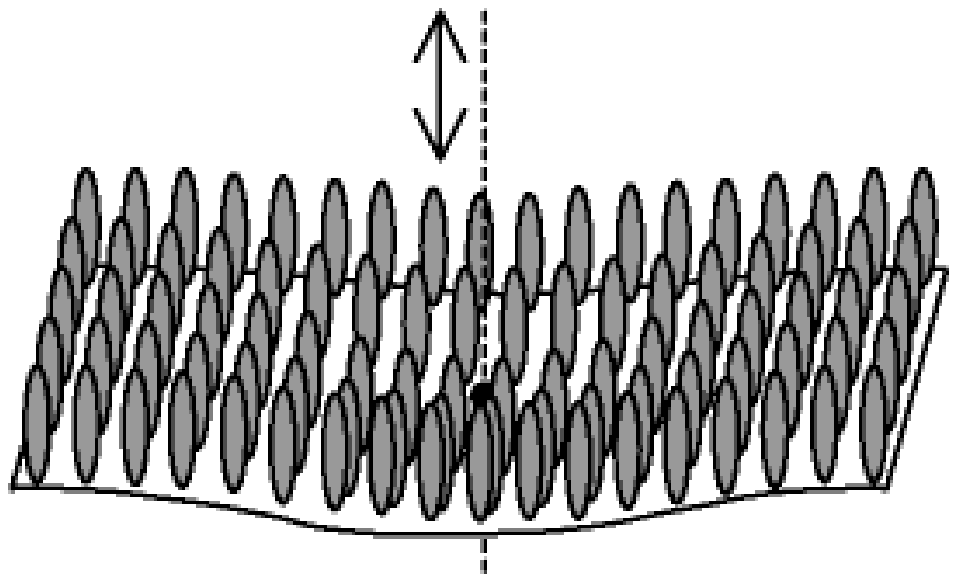}
  \end{tabular}
  \caption{Illustration of the oscillatory sources.
We consider the two geometries: (a) the one-dimensional wall and (b) the two-dimensional cylinder at the center point of the layer plane.}
  \label{fig:source}
 \end{center}
\end{figure}

We derive the equation of motion of $u(\br)$ from the time evolution of the velocity field $\bv(\br)$.
According to the unified hydrodynamic description~\cite{deGennes, MartinParodi},
a set of the hydrodynamic equations at a constant temperature with no external field is
\seq
\div{\bv(\br)}&=&0,\\
\der{u(\br)}{t}-v_z(\br)&=&0,\\
-B_{z\beta\gamma}\D_\gamma v_\beta(\br)&=&-\nu h_z(\br),\\
\rho_0\der{v_z(\br)}{t}&=&-\fder{F}{u}-B_{zz\beta}\D_\beta h_z(\br)+\eta_{z\beta\gamma\delta}\D_\gamma \D_\delta v_\beta(\br),
\sqe
where $h_z$ is the torque acting on the $\bc$-vector
and $B_{\alpha\beta\gamma}$
is the coupling constant between the torque and the velocity.
The first equation is the incompressibility condition originated from the equation of continuity.
The second and third equation describes the relaxation of the layer permeation and the rotational $\bc$-vector mode respectively.
The last equation is derived from the momentum conservation essentially equivalent to the Navier-Stokes equation.
From the above equations, we derive the equation of motion as
\eq
\rho_0 \der{^2 u(\br)}{t^2} &=& \mu_{\beta\gamma} \D_{\beta} \D_{\gamma} \der{u(\br)}{t} - \fder{F}{u}.
\label{eq:HY}
\qe
In the smectic-C phase, the model free energy has the static elastic coupling of the layer displacement field $u(\br)$ and $\bc$-vector.
Here exists the dynamical viscous coupling of the velocity and the molecular orientation in the first term of \Eq{HY}.
Anisotropic viscosity tensor $\mu_{\beta\gamma}$ ($\beta$, $\gamma$=1, 2) depends on the $\bc$-vector, as
\eq
\mu_{\beta\gamma} &=& \mu_0 \, \delta_{\beta\gamma} + \mu_1 n_\beta(\br) n_\gamma(\br),
\qe
where $\mu_0$ and $\mu_1$ are constants.
The isotropic and rotational viscosities are reported to be of the same order ($\sim 1$ Poise) in the experimental studies~\cite{Tamamushi, Meiboom}.
We then add an oscillatory source term $s(\br, t)$.
We consider the oscillating wall and cylindrical sources
$s(\br, t)=s_0 \cos(\omega t) \delta(x)$ and $s_0 \cos(\omega t) \delta(x)\delta(y)$ respectively,
to examine the anisotropy in the Sm-C phase (\Fig{source}).
This vibrating force can be provided by an oscillatory object, for example, an electric field~\cite{Pieranski}.
The linearized dynamic equation in Fourier space is
\eq
\rho_0 \der{^2 u(\bq, t)}{t^2} &=&
-\lrL{ (\mu_0 \, \delta_{\beta\gamma} + \mu_1 N_\beta N_\gamma) q_\beta q_\gamma \der{}{t} + \lrS{Bq_{\para}^2+K(\vartheta)q_{\perp}^4} } u(\bq, t) + s(\bq, t),
\label{eq:FHY}
\qe
where
$s(\bq, t)=4\pi^2 s_0 \cos(\omega t) \delta^{(2)}(\bqp)$ and $2\pi s_0 \cos(\omega t) \delta(q_z)$
is the one-dimensional wall and two-dimensional cylindrical source respectively.
The thermodynamic force $\delta F/\delta u^*$
is written in the linear form $L(\bq)u(\bq, t)=\lrS{Bq_{\para}^2+K(\vartheta, \bqp)q_{\perp}^4}u(\bq, t)$.
\par
We next derive the dimensionless dynamic equation.
The characteristic parameter set is listed in \Tab{param}.
\begin{table}[htb]
\caption{Parameters and their typical scales in the dynamic equation \Eq{FHY}~\cite{deGennes, Tamamushi, ChenJasnow, Meiboom}.}
\begin{tabular*}{16cm}{@{\extracolsep{\fill}}l l l}
\hline
\hline
parameter											&	description	&	typical scale\\
\hline
$\rho_0$											&	averaged molecular density				&	$\sim 1$g$/$cm$^3$\\
$u(\br)$											&	layer displacement field				&	$\sim 10^{-7}-10^{-8}$cm\\
$\mu_0$												&	isotropic viscosity						&	$\sim 1$ Poise\\
$\mu_1$												&	rotational viscosity					&	$\sim 1$ Poise\\
$q_i$												&	undulation wave number					&	$\sim 10^{-5}$cm\\
$\omega$											&	undulation frequency					&	$\sim 10^4$Hz\\
$B=C_i\Psi_0^2 q_0^2\,(i=\perp, \para)$				&	layer compression elastic constant		&	$\sim 10^{6}$dyn$/$cm$^2$\\
													&	(second order density elastic constant)	&\\
$D_i\Psi_0^2 q_0^2\,(i=\perp, \perp\para, \para)$	&	layer bending elastic constant			&	$\sim 10^{-6}$dyn\\
													&	(fourth order density elastic constant)	&\\
$K_i\,(i=1,2,3)$									&	Frank elastic modulus					&	$\sim 10^{-6}$dyn\\
\hline
\hline
\end{tabular*}
\label{tab:param}
\end{table}
The reduced dynamic equation is
\eq
\f{\rho_0 \xi^4}{K_3 \tau^2} \f{\D^2 \t{u}(\t{\bq}, \t{t})}{\D \t{t}^2}&=&
-\f{\xi^2\eta_c}{K_3 \tau}
\lrS{ \t{\eta}_0\t{q}^2 + \t{\eta}_1\lrS{\bN\cdot\t{\bq}}^2 }
\f{\D\t{u}(\t{\bq}, \t{t})}{\D\t{t}}\nn\\
&&-\lrS{\f{B_0}{K_3 q_0^2}\t{B}\t{q}_z^2
+\t{K}(\vartheta)\t{q}^4} \t{u}(\t{\bq}, \t{t}) + \f{\xi^4 q_0}{K_3}s(\t{\bq}, \t{t}),
\label{eq:RFHY}
\qe
where we introduced the dimensionless valuables
$t=\tau\t{t}$, $\omega\tau=\t\omega$, $u q_0=\t{u}$,
$q_i\xi=\t{q}_i$, $\eta_i=\eta_c \t{\eta}_i \,\, (i=0,1)$,
$B=B_0\t{B}$ and $K(\vartheta)=K_3\t{K}(\vartheta)$.
All the dimensionless constants are of order $\sim 1$.
We set the characteristic scales according to \Tab{param}:
$\tau=10^{-4}$s, $\xi=10^{-5}$cm, $\eta_c=1$ Poise and $B_0=10^6$dyn$/$cm$^2$.
In this parameter set, the scale of each term in \Eq{RFHY} is
$\rho_0 \xi^4/K_3 \tau^2 \sim 10^{-6}$ and $\xi^2\eta_c/K_3 \tau \sim B_0/K_3 q_0^2 \sim 1$.
Thus the inertia term can be neglected and the final form of the dimensionless dynamic equation is
\eq
\t{\zeta}\f{\D \t{u}(\t{\bq}, \t{t})}{\D\t{t}}+\t{L}(\t{\bq})\t{u}(\t{\bq}, \t{t})=\t{s}(\t{\bq}, \t{t}),
\label{eq:RHY}
\qe
where the effective viscosity $\t\zeta=\t{q}^2\lrS{\t\eta_0+\t\eta_1\cos^2\vartheta}$,
and the normalized external force $\t{s}$ is given by $\t{s}=\t{s}_0(\t\bq)\cos(\t{\omega}\t{t})$.
The reduced amplitude $\t{s}_0$ is $4\pi^2 s_0 q_0 \xi^4 \delta(q_y)\delta(q_z)/K_3$ and $2\pi s_0 q_0 \xi^4 \delta(q_z)/K_3$
in the oscillatory wall and the vibrating cylinder case respectively.
The time evolution is
\eq
\t u (\t{\bq}, \t{t}) =
\f{\t{s}_0}{\t\zeta\lrS{\t\lambda^2+\t\omega^2}}
\lrS{ -\t\lambda \cos\t\omega\t{t} + \t\omega\sin\t\omega\t{t}+\t\lambda\exp\t\lambda\t{t} },
\qe
where $\t\lambda=-\t L/\t\zeta$ and the initial condition is set to be $\t u(\t\bq, 0)=0$.
\par
\begin{figure}[htbp]
 \leavevmode
  \begin{center}
  \begin{tabular}{ clclcl }
	&& (a) $t=1.2\times10^{-4}$s
	& (b) $t=1.8\times10^{-4}$s\\
	\includegraphics[width=25mm]{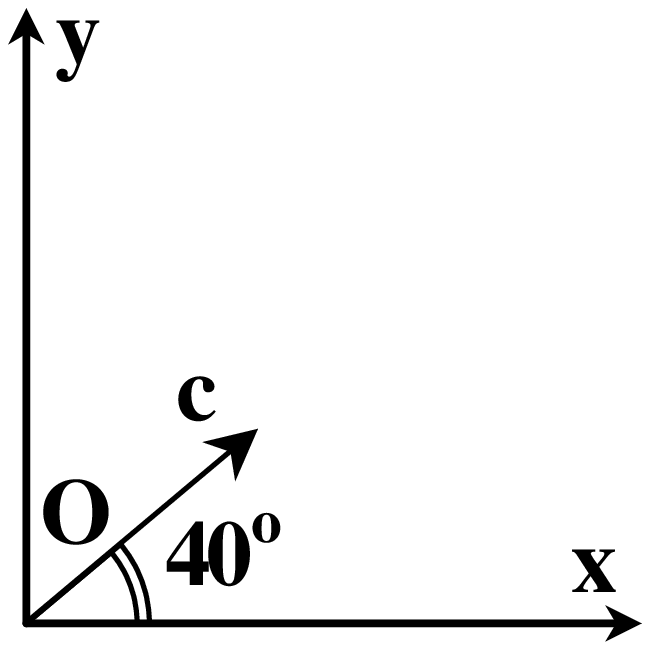}
	&&
	\includegraphics[width=28mm]{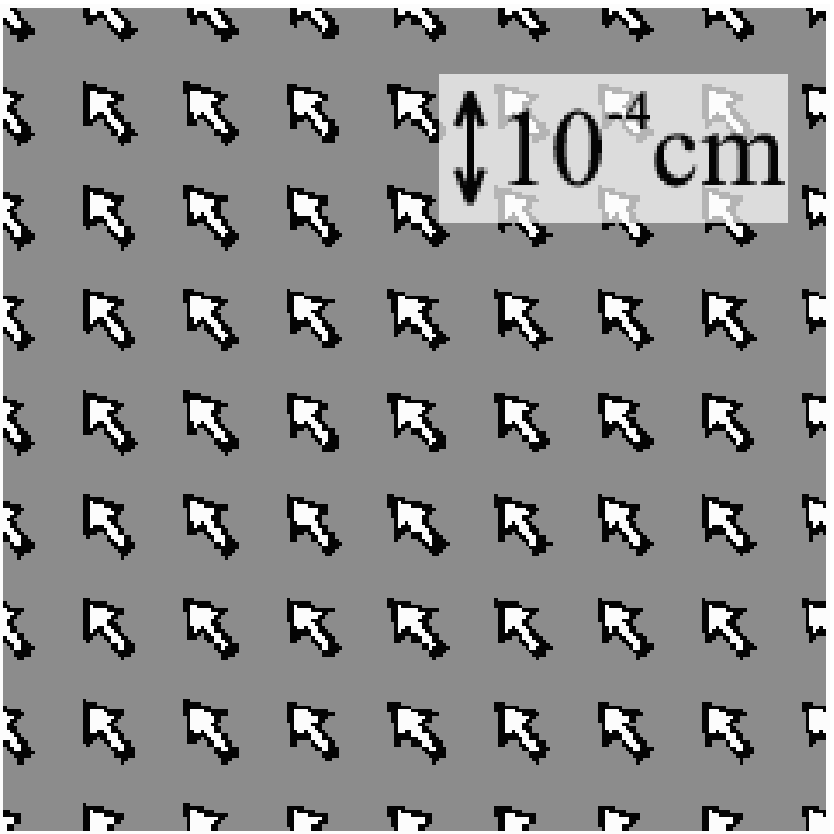}
	&
	\includegraphics[width=28mm]{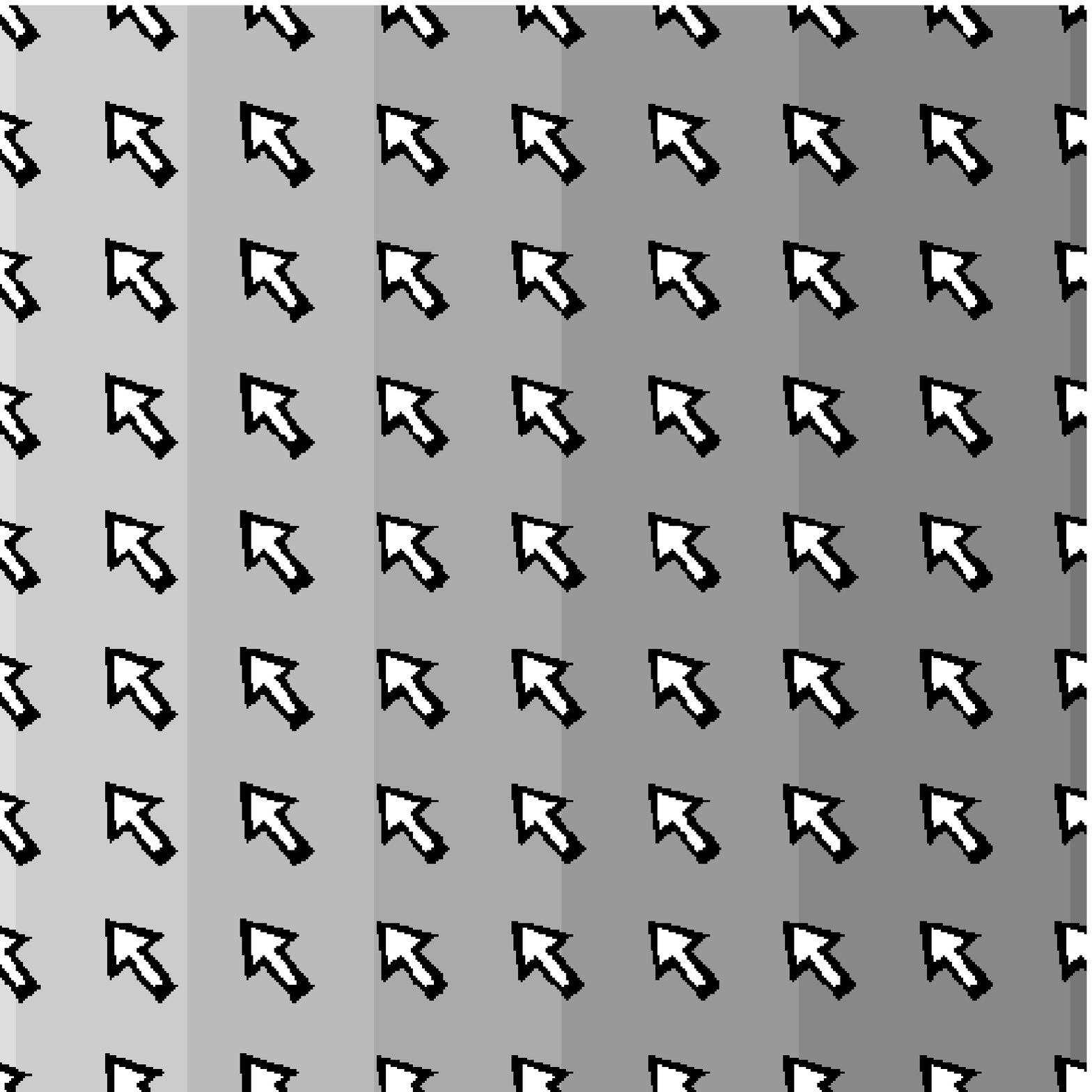}\\
	&& (c) $t=2.4\times10^{-4}$s
	& (d) $t=3.0\times10^{-4}$s\\
	\includegraphics[width=25mm]{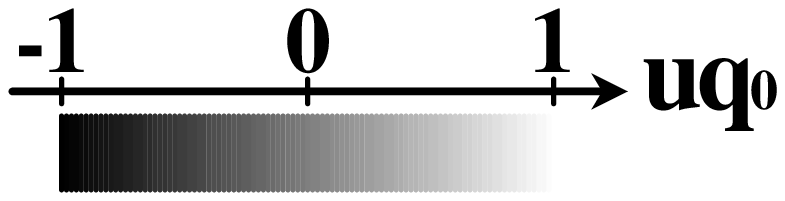}
	&&
	\includegraphics[width=28mm]{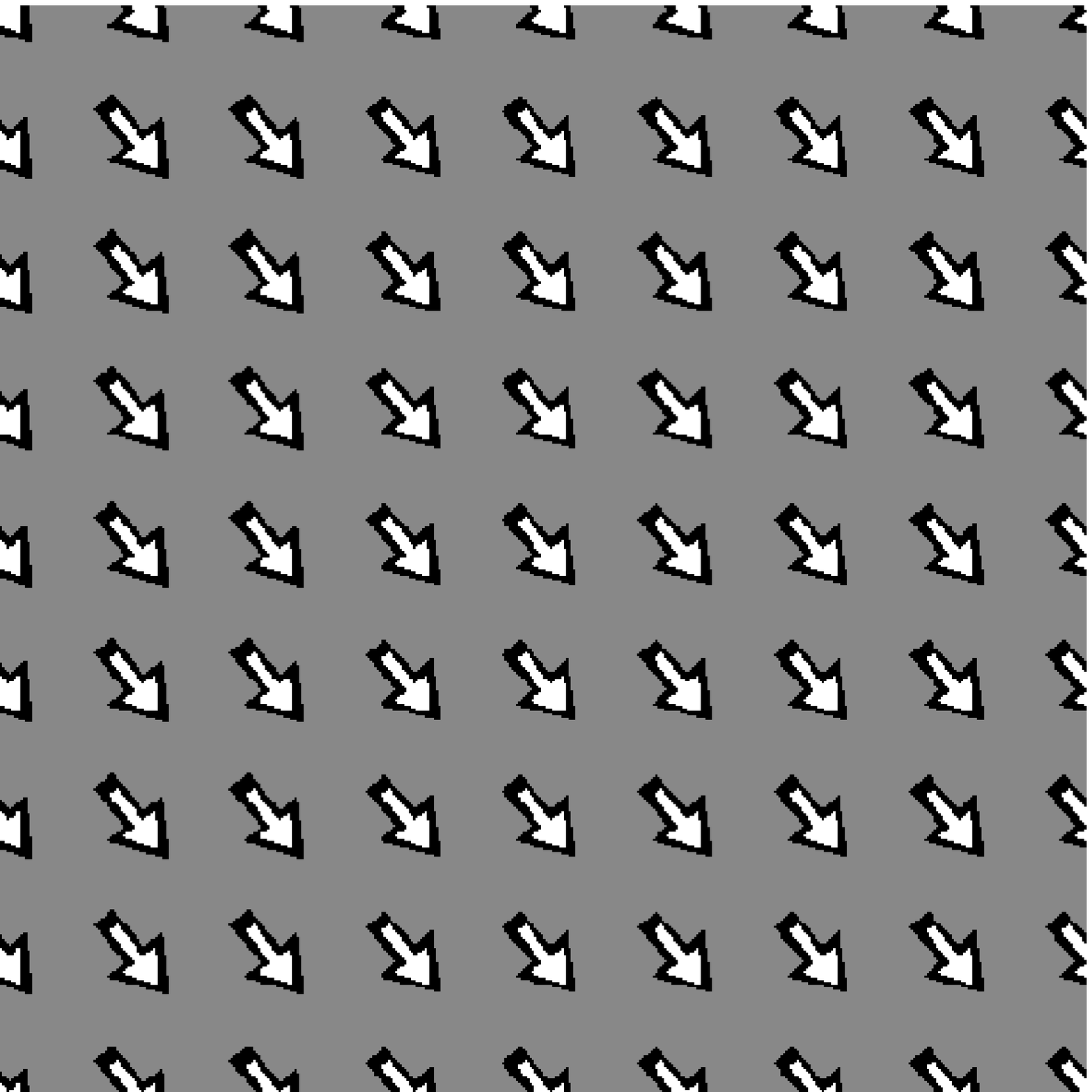}
	&
	\includegraphics[width=28mm]{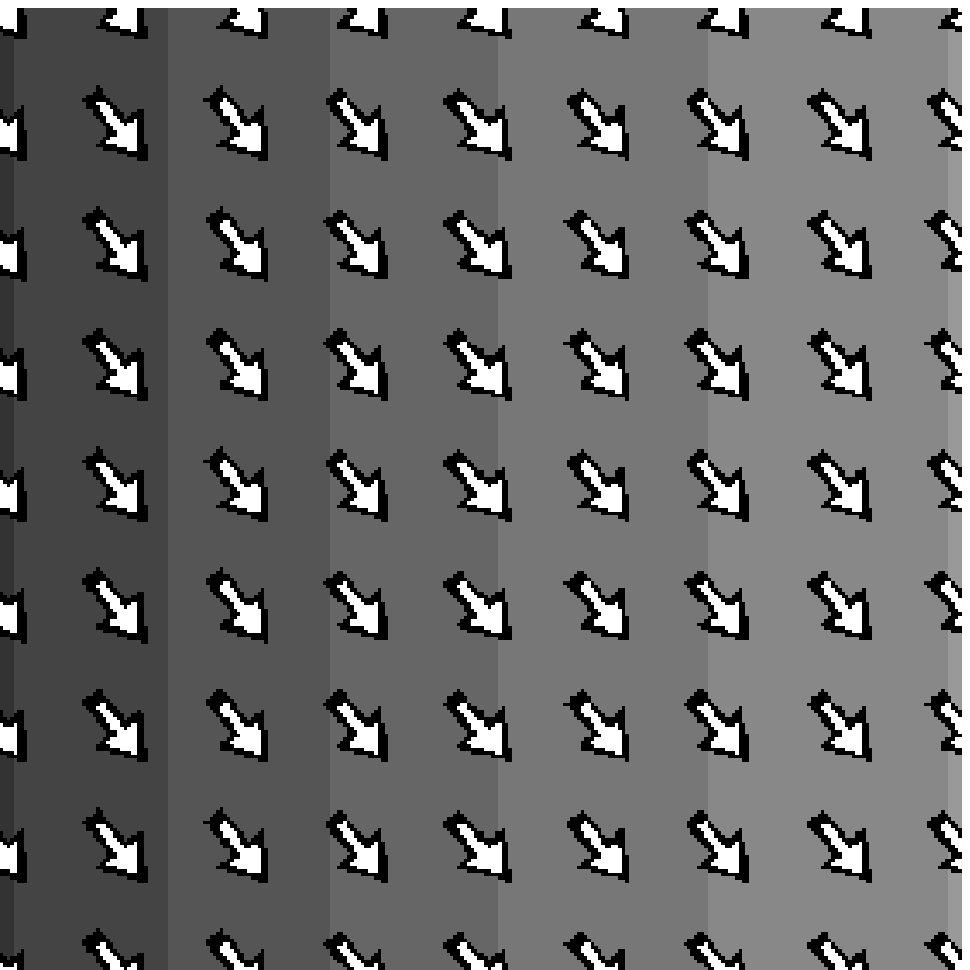}
  \end{tabular}
  \caption{Snapshots of the layer oscillation under a planar source ($x=0$).
The brightness indicates the local layer displacement and the unit arrow represents the deviation of the director from the ground state.
We set $\Cpl=4\times10^{12}$cm$^{-2}$, $\Dpr=\Dpp=\Dpl=0.1$,
$\tk1=\k2=0$, $\omega=5\times 10^4$Hz, and $\alpha=30\deg$.}
  \label{fig:pla}
 \end{center}
\end{figure}
\begin{figure}[htbp]
 \leavevmode
  \begin{center}
  \begin{tabular}{ clclcl }
	&& (a) $t=1.2\times10^{-4}$s
	& (b) $t=1.8\times10^{-4}$s\\
	\includegraphics[width=25mm]{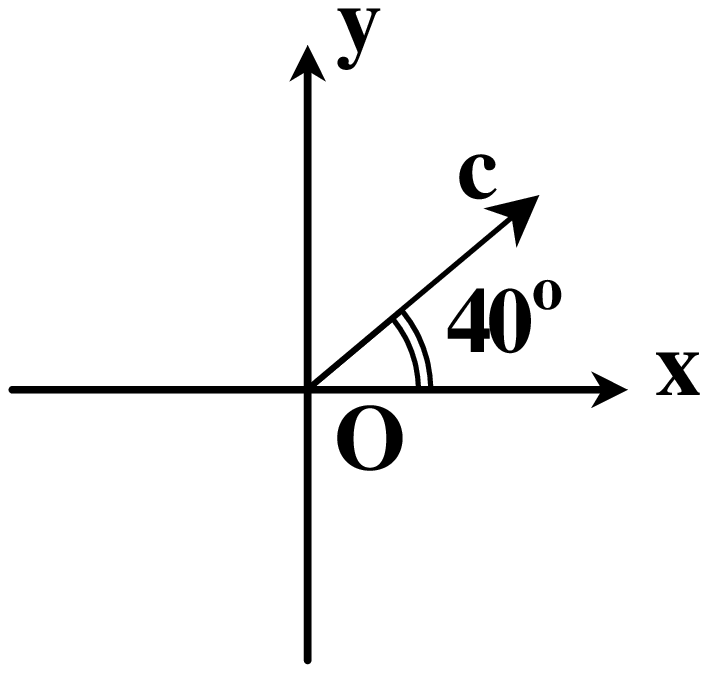}
	&&
	\includegraphics[width=28mm]{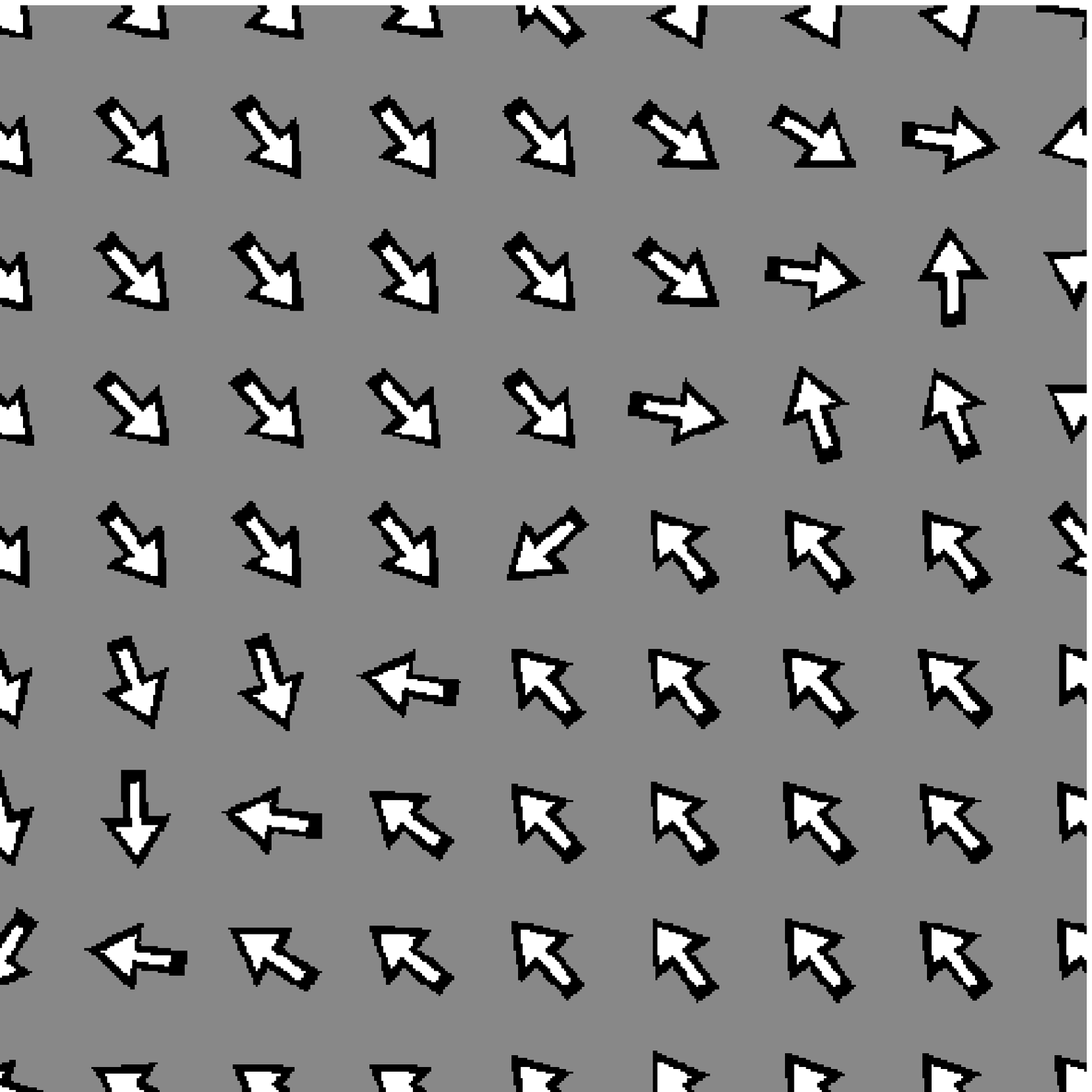}
	&
	\includegraphics[width=28mm]{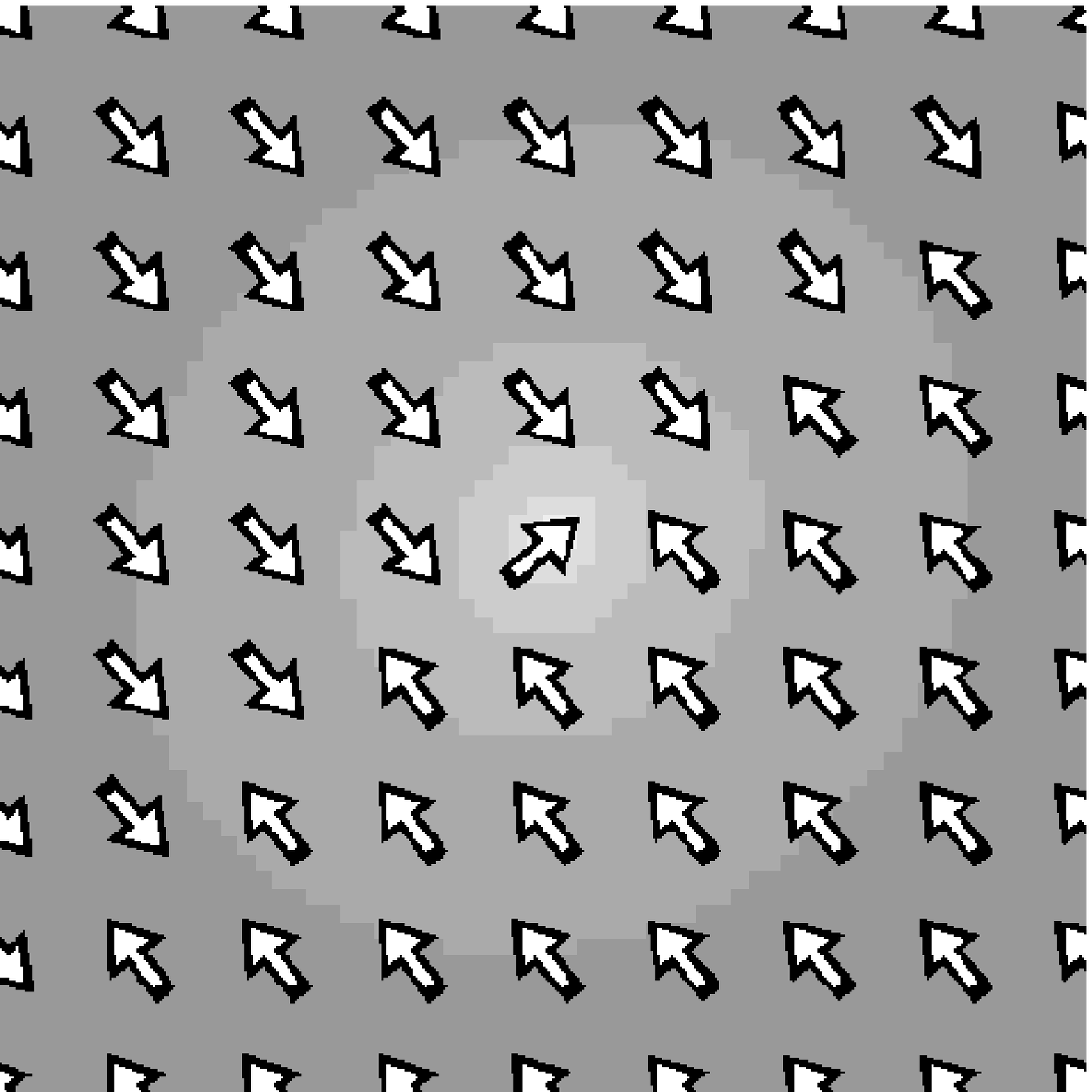}\\
	&& (c) $t=2.4\times10^{-4}$s
	& (d) $t=3.0\times10^{-4}$s\\
	&&
	\includegraphics[width=28mm]{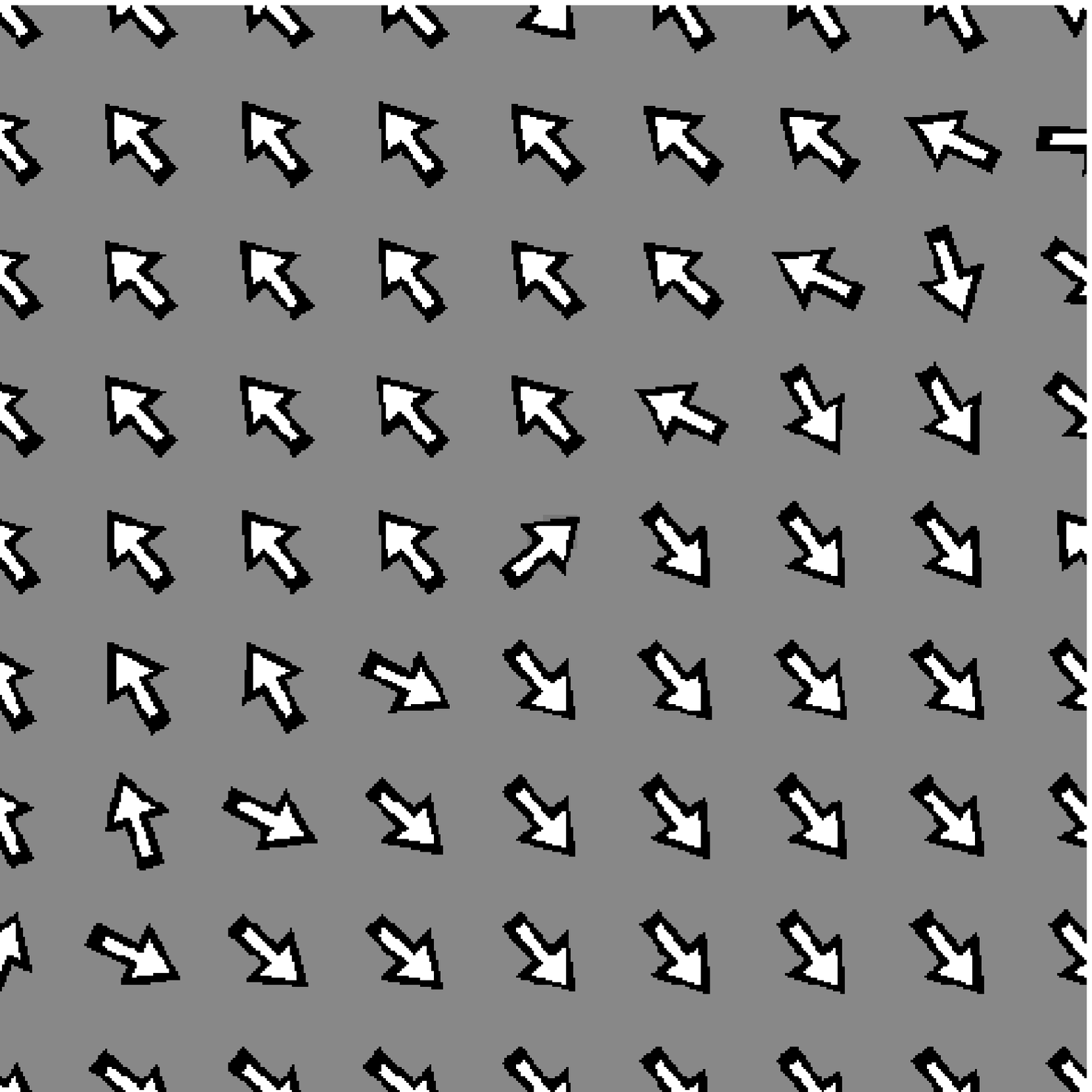}
	&
	\includegraphics[width=28mm]{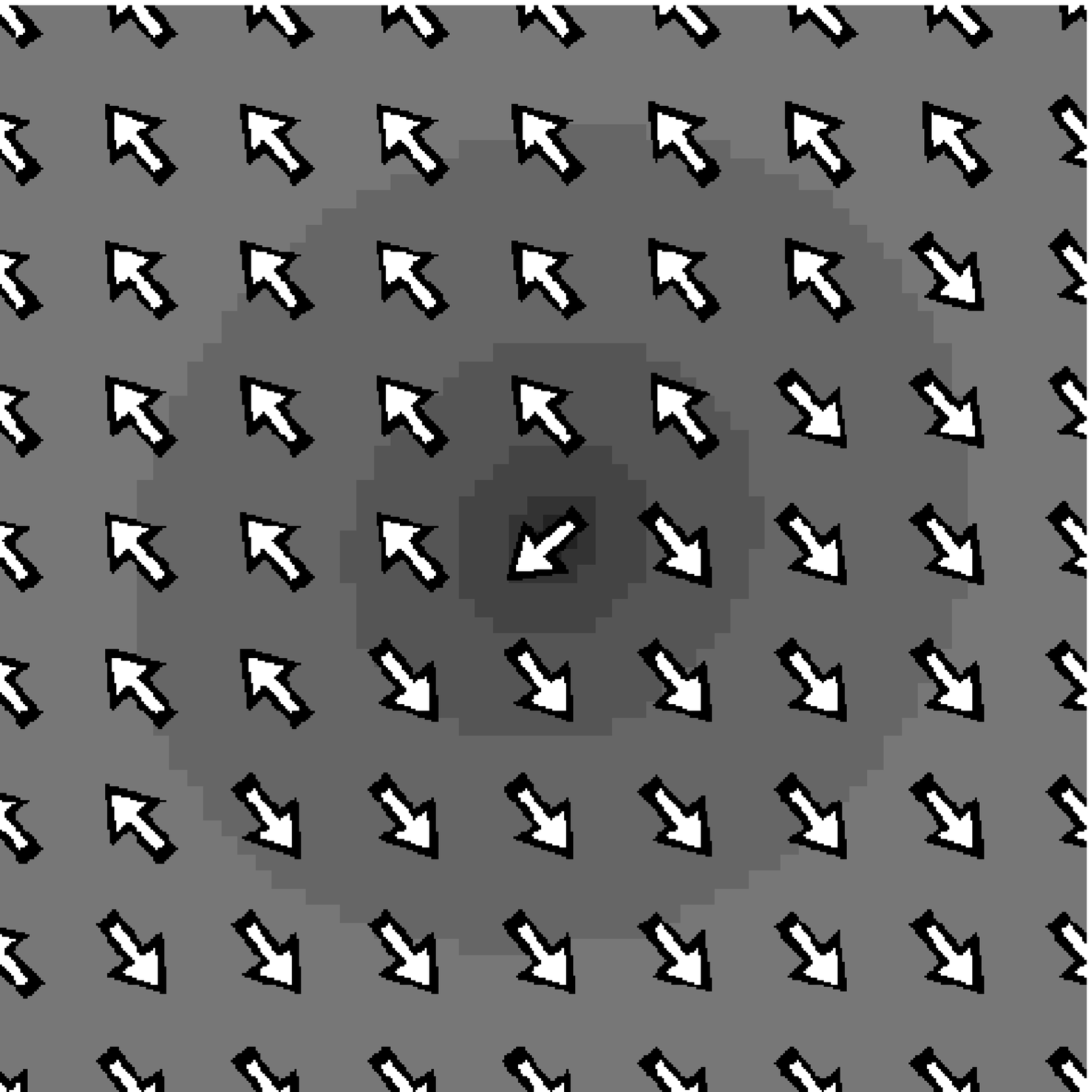}
  \end{tabular}
  \caption{Anisotropic wave propagation is shown in the presence of an isotropic oscillatory source ($x=y=0$).
The wave front is elliptic with the long axis parallel to the reference director $\bc$.}
  \label{fig:cyl}
 \end{center}
\end{figure}

Visualization of the wave propagation is depicted in \Fig{pla} and \Fig{cyl}.
Unphysical infinite wavelength mode is excluded.
We use the parameter set:
$\Cpl=4\times10^{12}$cm$^{-2}$, $\Dpr=\Dpp=\Dpl=0.1$,
$\tk1=\k2=0$, $\omega=5\times 10^4$Hz, and $\alpha=30\deg$.
The reference $\bc$-vector is tilted against the $x$-axis at $40\deg$.
The same parameter set is used in the following if not specified.
The layer displacement is shown with the brightness,
and the director change is indicated with the unit arrow.
The dumped director degree of freedom is governed by the layer displacement field through \Eq{CLCn}.
While the layer displacement scale is $\sim10^{-7}$cm$\inv$, the director rotation angle is about $10^{-3}-10^{-4}$rad,
which can be observed in experiment~\cite{Galerne, Kuo}.
\par
In \Fig{pla}, the planar oscillation orientates the director with the layer displacement.
Anisotropic wave propagation is obvious under a cylindrical source (\Fig{cyl}).
The favored angle is $\theta=90\deg$,
as in the experimental study~\cite{Johnson}
and the free energy analysis (\Fig{1FSDA}).
The wave velocity is faster in the rigid direction $\bc\para\bqp$,
than in the soft orientation $\bc\perp\bqp$.
A simple dimensional analysis of \Eq{RHY} with $\t{s}=0$ gives
the ratio
\eq
v_{\para}/v_{\perp}=\lrS{\t{K}(0\deg)/\t{K}(90\deg)}^{\f{1}{2}} \lrM{\t{\eta}_0/\lrS{\t{\eta}_0+\t{\eta}_1}}^{\f{1}{2}}.
\label{eq:vratio}
\qe

%
\par
We conclude this section with the remark that both the director alignment and the anisotropic wave propagation
are entirely controlled by the static and dynamic coupling between the layer displacement and the director \Eq{vratio},
and could be observed within the experimental resolution.
%
%
\section{Conclusion}
In this paper,
we analytically derived the effective generalized Chen-Lubensky model by adiabatic elimination of the director relaxation.
For Sm-A phase, the director unlocking from the layer normal is confirmed
at an undulation wavelength shorter than the director coherent length.
The effective layer bending elastic modulus is written as a function of the unlocking angle $\alpha$,
and decays with the increase of $\alpha$.
This agrees with the argument in the previous work~\cite{OgawaTGB}.
In Sm-C phase, on the other hand, not only the director unlocking
but anisotropic elasticity arises from the $\bc$-vector.
After the detailed study,
it turned out that the preferred angle $\theta$ between the layer bending orientation and the $\bc$-vector 
varies from $0\deg$ to $90\deg$ depending on the layer elastic constants, the Frank constants, the tilt angle,
and the undulation wavelength.
Transitions between the different $\theta$ states are not only continuous but also discontinuous.
Then the new characteristic length $\lambda_A$ is found,
which plays an important role on the elasticity of Sm-C phase.
The model parameters of the macroscopic free energy ~\cite{Hatwalne} are determined from
the more microscopic Chen-Lubensky model.
It allows a quantitative argument based on the model ~\cite{Hatwalne}
with the experimental determination of the model parameters~\cite{Martinez}.
We also compared smectic liquid crystals with other layer forming materials.
\par
Next we discussed the hydrodynamics of Sm-C layers using the effective elastic energy
derived above.
We derive the Sm-C hydrodynamics with the director deformation eliminated.
Anisotropic wave propagation from the isotropic oscillatory source
and the director orientation in the simple undulation along one direction
are confirmed.
These effect could be observed in experiment ~\cite{Pieranski, Galerne, Kuo}.
By converting the mechanical undulation to the optical information through the director orientation,
here also arises a new possible application of liquid crystals
for a sensitive mechanical sensor.
%
%
\begin{acknowledgments}
The author thanks Nariya Uchida, Toshihiro Kawakatsu, Helmut Brand, Hiroaki Honda and Takashi Shibata
for valuable discussions, useful suggestions, insightful comments, and carefully reading the manuscript.
This work is supported financially by the twenty-first century COE program of Tohoku University.
\end{acknowledgments}

\appendix

\section{Effective layer bending elastic modulus as a function of $\bem$-$\bn$ unlocking angle}
\begin{figure}[htbp]
  \begin{center}
	\includegraphics[width=50mm]{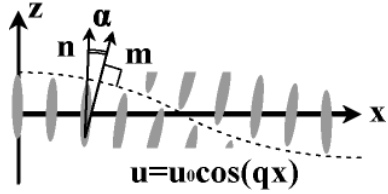}
  \caption{Schematic representation of the smectic layer with a sinusoidal undulation.
Here exists the small unlocking angle $\alpha$ between the layer normal $\bem$ and the director $\bn$,
which accounts for the weak bending elastic modulus.}
  \label{fig:Keff}
 \end{center}
\end{figure}
Here we give a better representation of the effective layer bending elastic modulus \Eq{Keffa},
as a function of the root mean square (RMS) of the angle $\talpha$ between the layer normal and the director.
This leads us to a more quantitative understanding
of the previous numerical result on the mean curvature of smectic layers in the TGB phase~\cite{OgawaTGB}.
\par
We suppose a single Fourier mode of the layer undulation $u(\br)=u_0\cos(qx)$ (\Fig{Keff}).
The wave-number vector is along the $x$-axis.
With the help of \Eq{L-deGnu}, $\bem$ and $\bn$ are
\seq
n_x &=& \f{1}{1+(\lambda q)^2} qu_0\sin(qx),\\
m_x &=& qu_0\sin(qx),\\
n_y &=& m_y=0,\\
n_z &=& m_z=1.
\sqe
In a small deformation, the tilt angle $\alpha$ is given by
\eq
\alpha &\simeq& m_x-n_x=\f{(\lambda q)^2}{1+(\lambda q)^2} qu_0\sin(qx).
\qe
The RMS $\talpha$ is readily calculated
\eq
\talpha=\f{1}{\sqrt{2}} \f{u_0\lambda^2  q^3}{1+(\lambda q)^2} \simeq \f{1}{\sqrt{2}} u_0\lambda^2  q^3,
\qe
where the smooth undulation condition $q \ll \lambda\inv$ is assumed in the last line.
We note that the tilt angle is roughly proportional to the cube of the wave-number,
and the characteristic length is a combination of the penetration length and the undulation amplitude $\lrS{\lambda^2 u_0}^{1/3}$.
Thus the effective layer bending rigidity is now written as a function of $\talpha$
\eq
\tilde{\kappa}\lrS{\talpha}=\f{1}{1+\lrS{\sqrt{2}\talpha\lambda/u_0}^{2/3}}.
\qe
The same method could be applied for the Sm-A phase also in the Chen-Lubensky model.
\section{Derivation of $\Dpr\Dpl-\Dpp^2>0$ and one constant approximation}
Here we prove the relation $\Dpr\Dpl-\Dpp^2>0$
to guarantee the convexity of the effective free energy
as a function of $x=\cos^2\vartheta$ in the one Frank constant case \Eq{fddx}.
This relation actually holds for an arbitrary set of Frank elastic moduli,
because the inequality is based on the stability of the smectic layer.
With the density profile $\Psi=\Psi_0\exp(\bq\cdot\br)$,
the $\bq$-dependent part of the Chen-Lubensky Hamiltonian \Eq{CL} is proportional to
\eq
\int d\br \Psi_0^2 \lrL{\Dpl Q_{\para}^4 + 2\Dpp Q_{\para}^2Q_{\perp}^2 + \Dpr Q_{\perp}^4},
\label{eq:FQ}
\qe
where the effective momenta $Q_{\para}^2=(q_{\para}-q_0)^2+q_{0\para}^2$
and $Q_{\perp}^2=q_{\perp}^2-q_{0\perp}^2$,
the characteristic wave number 
$q_{0\para}^2=\lrS{\Cpl\Dpr-\Cpr\Dpp} / \lrL{2\lrS{\Dpr\Dpl-2\Dpp}}$
and
$q_{0\perp}^2=\lrS{\Cpl\Dpp-\Cpr\Dpl}/\lrL{2\lrS{\Dpr\Dpl-2\Dpp^2}}$,
and the projected wave-number vector $\bq_{\para}=(\bn\cdot\bq)\bn$, $\bq_{\perp}=-\bn\times\lrS{\bn\times\bq}$ are utilized.
Positive definiteness of \Eq{FQ} requires
\eq
\Dpr\Dpl-\Dpp^2&>&0.
\label{eq:lstability}
\qe
This inequality helps one to prove a convexity of the free energy at the one constant approximation ($K_1=K_2=K_3$).
In this case, the free energy is simplified as
\eq
\f{\Dpr N_3^2}{\qp^4|u(\bq)|^2} \t{f} &=&
		2\Np^4\lrS{\Dpl\Dpr-\Dpp^2}x^2+\Dpr\lrS{1-\Np^2 x} + 2\Dpp \Np^2 x\nn\\
	&&	- \f{ \lrS{2\Np^2 + \kp^2}  \lrS{ -2\Np^4 \Dpp^2 x^2 + \Dpr\lrS{1-\Np^2 x} + 2\Dpp\Np^2 x} - 2 N_3^2\Np^2\Dpr x}
			{ 2 \kp^2  \Dpr\lrS{1-\Np^2 x} + 4 \Dpr\Np^2\lrS{1-x} + 2 \Np^2 + \kp^2 },\nn\\
\label{eq:1afx}
\qe
and the second derivative is
\eq
\f{N_3^2\Dpr}{4\Np^4\qp^4|u(\bq)|^2} \t{f}^{''} (x) &=& \Dpl\Dpr-\Dpp^2 + \f{ 2\Np^2+\kp^2 }
	{ \lrM{ \lrS{1+2\Dpr}\lrS{2\Np^2+\kp^2} - 2\Np^2\Dpr\lrS{\kp^2+2}x}^3 }\nn\\
	&&\times\lrM{\lrS{1+2\Dpr}\lrS{2\Np^2+\kp^2}\Dpp - \lrS{\kp^2+2}\Dpr}^2,
\label{eq:fddx}
\qe
where the dimensionless wave number is given by $\kp(\Dpp, \Dpr) \equiv \lambda_A\qp$.


%
\section{Splay Frank energy as a function of $\vartheta$}
We calculate the splay energy for a small tilt angle $\alpha$ and the angle $\vartheta$ between $\bc$ and $\bqp$.
Let the $x$-axis be parallel to
$\bqp$.
The director is $\bn=\lrS{\sin\alpha \cos\vartheta, \sin\alpha\sin\vartheta, \cos\alpha}$.
Assuming the smooth layer undulation $\qp \ll \lambda\inv$, the director deformation perfectly follows the layer normal vector.
Undulated director configuration is obtained by operating the rotation matrix about the $y$-axis $\bR_y$
\eq
\bR_y=
\left(
\begin{array}{c c c}
\cos qx	&	0	&	-\sin qx\\
0		&	1	&	0\\
\sin qx	&	0	&	\cos qx
\end{array}
\right),
\qe
to the reference director $\bn$. Thus the spatially averaged splay energy is expressed as
\eq
\lrA{\lrM{\div{\lrS{\bR_y \bn}}}^2} =\f{q^2}{2}\lrS{\sin^2\alpha \cos^2\vartheta + \cos^2\alpha}.
\qe
Stability of the $\theta=90\deg$ state grows as the splay Frank elasticity dominates.
However this effect of the splay term is not very strong, due to the factor $\sin^2\alpha$.
\bibliography{paper3.56}

\begin{thebibliography}{54}
\expandafter\ifx\csname natexlab\endcsname\relax\def\natexlab#1{#1}\fi
\expandafter\ifx\csname bibnamefont\endcsname\relax
  \def\bibnamefont#1{#1}\fi
\expandafter\ifx\csname bibfnamefont\endcsname\relax
  \def\bibfnamefont#1{#1}\fi
\expandafter\ifx\csname citenamefont\endcsname\relax
  \def\citenamefont#1{#1}\fi
\expandafter\ifx\csname url\endcsname\relax
  \def\url#1{\texttt{#1}}\fi
\expandafter\ifx\csname urlprefix\endcsname\relax\def\urlprefix{URL }\fi
\providecommand{\bibinfo}[2]{#2}
\providecommand{\eprint}[2][]{\url{#2}}

\bibitem[{\citenamefont{de~Gennes and Prost}(1994)}]{deGennes}
\bibinfo{author}{\bibfnamefont{P.~G.} \bibnamefont{de~Gennes}}
  \bibnamefont{and} \bibinfo{author}{\bibfnamefont{J.}~\bibnamefont{Prost}},
  \emph{\bibinfo{title}{The Physics of Liquid Crystals}}
  (\bibinfo{publisher}{Clarendon, Oxford}, \bibinfo{year}{1994}).

\bibitem[{\citenamefont{Chainkin and Lubensky}(1995)}]{Lubensky}
\bibinfo{author}{\bibfnamefont{P.~M.} \bibnamefont{Chainkin}} \bibnamefont{and}
  \bibinfo{author}{\bibfnamefont{T.~C.} \bibnamefont{Lubensky}},
  \emph{\bibinfo{title}{Principles of Condensed Matter Physics}}
  (\bibinfo{publisher}{Cambridge University Press, Cambridge, England},
  \bibinfo{year}{1995}).

\bibitem[{\citenamefont{Keyes et~al.}(1973)\citenamefont{Keyes, Weston, and
  Daniels}}]{Keyes}
\bibinfo{author}{\bibfnamefont{P.~H.} \bibnamefont{Keyes}},
  \bibinfo{author}{\bibfnamefont{H.~T.} \bibnamefont{Weston}},
  \bibnamefont{and} \bibinfo{author}{\bibfnamefont{W.~B.}
  \bibnamefont{Daniels}}, \bibinfo{journal}{Phys.\ Rev.\ Lett.}
  \textbf{\bibinfo{volume}{31}}, \bibinfo{pages}{628} (\bibinfo{year}{1973}).

\bibitem[{\citenamefont{Aharony et~al.}(1986)\citenamefont{Aharony, Birgeneau,
  Brock, and Litster}}]{Aharony}
\bibinfo{author}{\bibfnamefont{A.}~\bibnamefont{Aharony}},
  \bibinfo{author}{\bibfnamefont{R.~J.} \bibnamefont{Birgeneau}},
  \bibinfo{author}{\bibfnamefont{J.~D.} \bibnamefont{Brock}}, \bibnamefont{and}
  \bibinfo{author}{\bibfnamefont{J.~D.} \bibnamefont{Litster}},
  \bibinfo{journal}{Phys.\ Rev.\ Lett.} \textbf{\bibinfo{volume}{57}},
  \bibinfo{pages}{1012} (\bibinfo{year}{1986}).

\bibitem[{\citenamefont{Shashidhar et~al.}(1984)\citenamefont{Shashidhar,
  Ratna, and Prasad}}]{Shashidhar}
\bibinfo{author}{\bibfnamefont{R.}~\bibnamefont{Shashidhar}},
  \bibinfo{author}{\bibfnamefont{B.~R.} \bibnamefont{Ratna}}, \bibnamefont{and}
  \bibinfo{author}{\bibfnamefont{S.~K.} \bibnamefont{Prasad}},
  \bibinfo{journal}{Phys.\ Rev.\ Lett.} \textbf{\bibinfo{volume}{53}},
  \bibinfo{pages}{2141} (\bibinfo{year}{1984}).

\bibitem[{\citenamefont{Drossinos and Ronis}(1986)}]{Drossinos}
\bibinfo{author}{\bibfnamefont{Y.}~\bibnamefont{Drossinos}} \bibnamefont{and}
  \bibinfo{author}{\bibfnamefont{D.}~\bibnamefont{Ronis}},
  \bibinfo{journal}{Phys.\ Rev.\ A} \textbf{\bibinfo{volume}{33}},
  \bibinfo{pages}{589} (\bibinfo{year}{1986}).

\bibitem[{\citenamefont{de~Gennes}(1973)}]{deGennesSmC}
\bibinfo{author}{\bibfnamefont{P.~G.} \bibnamefont{de~Gennes}},
  \bibinfo{journal}{Mol.\ Cryst.\ Liq.\ Cryst.} \textbf{\bibinfo{volume}{21}},
  \bibinfo{pages}{49} (\bibinfo{year}{1973}).

\bibitem[{\citenamefont{Chu and McMillan}(1977)}]{Chu}
\bibinfo{author}{\bibfnamefont{K.~C.} \bibnamefont{Chu}} \bibnamefont{and}
  \bibinfo{author}{\bibfnamefont{W.~L.} \bibnamefont{McMillan}},
  \bibinfo{journal}{Phys.\ Rev.\ A} \textbf{\bibinfo{volume}{15}},
  \bibinfo{pages}{1181} (\bibinfo{year}{1977}).

\bibitem[{\citenamefont{Chen and Lubensky}(1976)}]{ChenLubensky}
\bibinfo{author}{\bibfnamefont{J.}~\bibnamefont{Chen}} \bibnamefont{and}
  \bibinfo{author}{\bibfnamefont{T.~C.} \bibnamefont{Lubensky}},
  \bibinfo{journal}{Phys.\ Rev.\ A} \textbf{\bibinfo{volume}{14}},
  \bibinfo{pages}{1202} (\bibinfo{year}{1976}).

\bibitem[{\citenamefont{Benguigui}(1979)}]{Benguigui}
\bibinfo{author}{\bibfnamefont{L.}~\bibnamefont{Benguigui}},
  \bibinfo{journal}{J.\ Phys.\ (Paris).\ Colloq.}
  \textbf{\bibinfo{volume}{40}}, \bibinfo{pages}{C3} (\bibinfo{year}{1979}).

\bibitem[{\citenamefont{Huang and Lien}(1981)}]{Huang}
\bibinfo{author}{\bibfnamefont{C.~C.} \bibnamefont{Huang}} \bibnamefont{and}
  \bibinfo{author}{\bibfnamefont{S.~C.} \bibnamefont{Lien}},
  \bibinfo{journal}{Phys.\ Rev.\ Lett.} \textbf{\bibinfo{volume}{47}},
  \bibinfo{pages}{1917} (\bibinfo{year}{1981}).

\bibitem[{\citenamefont{Grinstein and Toner}(1983)}]{Grinstein}
\bibinfo{author}{\bibfnamefont{G.}~\bibnamefont{Grinstein}} \bibnamefont{and}
  \bibinfo{author}{\bibfnamefont{J.}~\bibnamefont{Toner}},
  \bibinfo{journal}{Phys.\ Rev.\ Lett.} \textbf{\bibinfo{volume}{26}},
  \bibinfo{pages}{2386} (\bibinfo{year}{1983}).

\bibitem[{\citenamefont{Witanachchi et~al.}(1983)\citenamefont{Witanachchi,
  Huang, and Ho}}]{Witanachchi}
\bibinfo{author}{\bibfnamefont{S.}~\bibnamefont{Witanachchi}},
  \bibinfo{author}{\bibfnamefont{J.}~\bibnamefont{Huang}}, \bibnamefont{and}
  \bibinfo{author}{\bibfnamefont{J.~T.} \bibnamefont{Ho}},
  \bibinfo{journal}{Phys.\ Rev.\ Lett.} \textbf{\bibinfo{volume}{50}},
  \bibinfo{pages}{594} (\bibinfo{year}{1983}).

\bibitem[{\citenamefont{Safinya et~al.}(1983)\citenamefont{Safinya,
  Martinez-Miranda, Kaplan, Litster, and Birgeneau}}]{Safinya}
\bibinfo{author}{\bibfnamefont{C.~R.} \bibnamefont{Safinya}},
  \bibinfo{author}{\bibfnamefont{L.~J.} \bibnamefont{Martinez-Miranda}},
  \bibinfo{author}{\bibfnamefont{M.}~\bibnamefont{Kaplan}},
  \bibinfo{author}{\bibfnamefont{J.~D.} \bibnamefont{Litster}},
  \bibnamefont{and} \bibinfo{author}{\bibfnamefont{R.~J.}
  \bibnamefont{Birgeneau}}, \bibinfo{journal}{Phys.\ Rev.\ Lett.}
  \textbf{\bibinfo{volume}{50}}, \bibinfo{pages}{56} (\bibinfo{year}{1983}).

\bibitem[{\citenamefont{Martinez-Miranda
  et~al.}(1986)\citenamefont{Martinez-Miranda, Kortan, and
  Birgeneau}}]{Martinez}
\bibinfo{author}{\bibfnamefont{L.~J.} \bibnamefont{Martinez-Miranda}},
  \bibinfo{author}{\bibfnamefont{A.~R.} \bibnamefont{Kortan}},
  \bibnamefont{and} \bibinfo{author}{\bibfnamefont{R.~J.}
  \bibnamefont{Birgeneau}}, \bibinfo{journal}{Phys.\ Rev.\ Lett.}
  \textbf{\bibinfo{volume}{56}}, \bibinfo{pages}{2264} (\bibinfo{year}{1986}).

\bibitem[{\citenamefont{Kitzerow and Bahr}(2002)}]{Kitzerow}
\bibinfo{editor}{\bibfnamefont{H.~S.} \bibnamefont{Kitzerow}} \bibnamefont{and}
  \bibinfo{editor}{\bibfnamefont{C.}~\bibnamefont{Bahr}}, eds.,
  \emph{\bibinfo{title}{Chirality in Liquid Crystals}}
  (\bibinfo{publisher}{Springer-Verlag, New York}, \bibinfo{year}{2002}).

\bibitem[{\citenamefont{de~Gennes}(1972)}]{deGennesAna}
\bibinfo{author}{\bibfnamefont{P.~G.} \bibnamefont{de~Gennes}},
  \bibinfo{journal}{Solid State Commun.} \textbf{\bibinfo{volume}{10}},
  \bibinfo{pages}{753} (\bibinfo{year}{1972}).

\bibitem[{\citenamefont{Renn and Lubensky}(1988)}]{Renn}
\bibinfo{author}{\bibfnamefont{S.~R.} \bibnamefont{Renn}} \bibnamefont{and}
  \bibinfo{author}{\bibfnamefont{T.~C.} \bibnamefont{Lubensky}},
  \bibinfo{journal}{Phys.\ Rev.\ A} \textbf{\bibinfo{volume}{38}},
  \bibinfo{pages}{2132} (\bibinfo{year}{1988}).

\bibitem[{\citenamefont{Goodby et~al.}(1989)\citenamefont{Goodby, Waugh, Stein,
  Chin, Pindak, and Patel}}]{Goodby}
\bibinfo{author}{\bibfnamefont{J.~W.} \bibnamefont{Goodby}},
  \bibinfo{author}{\bibfnamefont{M.~A.} \bibnamefont{Waugh}},
  \bibinfo{author}{\bibfnamefont{S.~M.} \bibnamefont{Stein}},
  \bibinfo{author}{\bibfnamefont{E.}~\bibnamefont{Chin}},
  \bibinfo{author}{\bibfnamefont{R.}~\bibnamefont{Pindak}}, \bibnamefont{and}
  \bibinfo{author}{\bibfnamefont{J.~S.} \bibnamefont{Patel}},
  \bibinfo{journal}{Nature} \textbf{\bibinfo{volume}{337}},
  \bibinfo{pages}{449} (\bibinfo{year}{1989}).

\bibitem[{\citenamefont{Lubensky and Renn}(1990)}]{LubenskyRenn}
\bibinfo{author}{\bibfnamefont{T.~C.} \bibnamefont{Lubensky}} \bibnamefont{and}
  \bibinfo{author}{\bibfnamefont{S.~R.} \bibnamefont{Renn}},
  \bibinfo{journal}{Phys.\ Rev.\ A} \textbf{\bibinfo{volume}{41}},
  \bibinfo{pages}{4392} (\bibinfo{year}{1990}).

\bibitem[{\citenamefont{Ismaili et~al.}(2000)\citenamefont{Ismaili, Anakkar,
  Joly, and Isaert}}]{Ismaili}
\bibinfo{author}{\bibfnamefont{M.}~\bibnamefont{Ismaili}},
  \bibinfo{author}{\bibfnamefont{A.}~\bibnamefont{Anakkar}},
  \bibinfo{author}{\bibfnamefont{G.}~\bibnamefont{Joly}}, \bibnamefont{and}
  \bibinfo{author}{\bibfnamefont{N.}~\bibnamefont{Isaert}},
  \bibinfo{journal}{Phys.\ Rev.\ E} \textbf{\bibinfo{volume}{61}},
  \bibinfo{pages}{519} (\bibinfo{year}{2000}).

\bibitem[{\citenamefont{Grelet et~al.}(2001{\natexlab{a}})\citenamefont{Grelet,
  Pansu, Li, and Nguyen}}]{Grelet}
\bibinfo{author}{\bibfnamefont{E.}~\bibnamefont{Grelet}},
  \bibinfo{author}{\bibfnamefont{B.}~\bibnamefont{Pansu}},
  \bibinfo{author}{\bibfnamefont{M.}~\bibnamefont{Li}}, \bibnamefont{and}
  \bibinfo{author}{\bibfnamefont{H.~T.} \bibnamefont{Nguyen}},
  \bibinfo{journal}{Phys.\ Rev.\ Lett.} \textbf{\bibinfo{volume}{86}},
  \bibinfo{pages}{3791} (\bibinfo{year}{2001}{\natexlab{a}}).

\bibitem[{\citenamefont{DiDonna and Kamien}(2002)}]{DiDonna}
\bibinfo{author}{\bibfnamefont{B.~A.} \bibnamefont{DiDonna}} \bibnamefont{and}
  \bibinfo{author}{\bibfnamefont{R.~D.} \bibnamefont{Kamien}},
  \bibinfo{journal}{Phys.\ Rev.\ Lett.} \textbf{\bibinfo{volume}{89}},
  \bibinfo{pages}{215504} (\bibinfo{year}{2002}).

\bibitem[{\citenamefont{Yamamoto et~al.}(2005)\citenamefont{Yamamoto,
  Nishiyama, Inoue, and Yokoyama}}]{Yamamoto}
\bibinfo{author}{\bibfnamefont{J.}~\bibnamefont{Yamamoto}},
  \bibinfo{author}{\bibfnamefont{I.}~\bibnamefont{Nishiyama}},
  \bibinfo{author}{\bibfnamefont{M.}~\bibnamefont{Inoue}}, \bibnamefont{and}
  \bibinfo{author}{\bibfnamefont{H.}~\bibnamefont{Yokoyama}},
  \bibinfo{journal}{Nature} \textbf{\bibinfo{volume}{437}},
  \bibinfo{pages}{525} (\bibinfo{year}{2005}).

\bibitem[{\citenamefont{Ogawa and Uchida}(2006)}]{OgawaTGB}
\bibinfo{author}{\bibfnamefont{H.}~\bibnamefont{Ogawa}} \bibnamefont{and}
  \bibinfo{author}{\bibfnamefont{N.}~\bibnamefont{Uchida}},
  \bibinfo{journal}{Phys.\ Rev.\ E} \textbf{\bibinfo{volume}{73}},
  \bibinfo{pages}{060701(R)} (\bibinfo{year}{2006}).

\bibitem[{\citenamefont{Grelet et~al.}(2001{\natexlab{b}})\citenamefont{Grelet,
  Pansu, and Nguyen}}]{GreletTilt}
\bibinfo{author}{\bibfnamefont{E.}~\bibnamefont{Grelet}},
  \bibinfo{author}{\bibfnamefont{B.}~\bibnamefont{Pansu}}, \bibnamefont{and}
  \bibinfo{author}{\bibfnamefont{H.~T.} \bibnamefont{Nguyen}},
  \bibinfo{journal}{Phys.\ Rev.\ E} \textbf{\bibinfo{volume}{64}},
  \bibinfo{pages}{010703(R)} (\bibinfo{year}{2001}{\natexlab{b}}).

\bibitem[{\citenamefont{Ohta and Kawasaki}(1986)}]{OhtaKawasaki}
\bibinfo{author}{\bibfnamefont{T.}~\bibnamefont{Ohta}} \bibnamefont{and}
  \bibinfo{author}{\bibfnamefont{K.}~\bibnamefont{Kawasaki}},
  \bibinfo{journal}{Macromol.} \textbf{\bibinfo{volume}{19}},
  \bibinfo{pages}{2621} (\bibinfo{year}{1986}).

\bibitem[{\citenamefont{Gompper and Klein}(1992)}]{GompperKlein}
\bibinfo{author}{\bibfnamefont{G.}~\bibnamefont{Gompper}} \bibnamefont{and}
  \bibinfo{author}{\bibfnamefont{S.}~\bibnamefont{Klein}},
  \bibinfo{journal}{J.\ Phys. II (France)} \textbf{\bibinfo{volume}{2}},
  \bibinfo{pages}{1725} (\bibinfo{year}{1992}).

\bibitem[{\citenamefont{Leibler}(1980)}]{Leibler}
\bibinfo{author}{\bibfnamefont{L.}~\bibnamefont{Leibler}},
  \bibinfo{journal}{Macromol.} \textbf{\bibinfo{volume}{13}},
  \bibinfo{pages}{1602} (\bibinfo{year}{1980}).

\bibitem[{\citenamefont{Hatwalne and Lybensky}(1995)}]{Hatwalne}
\bibinfo{author}{\bibfnamefont{Y.}~\bibnamefont{Hatwalne}} \bibnamefont{and}
  \bibinfo{author}{\bibfnamefont{T.~C.} \bibnamefont{Lybensky}},
  \bibinfo{journal}{Phys.\ Rev.\ E} \textbf{\bibinfo{volume}{52}},
  \bibinfo{pages}{6240} (\bibinfo{year}{1995}).

\bibitem[{\citenamefont{Martin et~al.}(1972)\citenamefont{Martin, Parodi, and
  Pershan}}]{MartinParodi}
\bibinfo{author}{\bibfnamefont{P.~C.} \bibnamefont{Martin}},
  \bibinfo{author}{\bibfnamefont{O.}~\bibnamefont{Parodi}}, \bibnamefont{and}
  \bibinfo{author}{\bibfnamefont{P.~S.} \bibnamefont{Pershan}},
  \bibinfo{journal}{Phys.\ Rev.\ A} \textbf{\bibinfo{volume}{6}},
  \bibinfo{pages}{2401} (\bibinfo{year}{1972}).

\bibitem[{\citenamefont{Buka and Kramer}(1996)}]{Buka}
\bibinfo{editor}{\bibfnamefont{A.}~\bibnamefont{Buka}} \bibnamefont{and}
  \bibinfo{editor}{\bibfnamefont{L.}~\bibnamefont{Kramer}}, eds.,
  \emph{\bibinfo{title}{Pattern Formation in Liquid Crystals}}
  (\bibinfo{publisher}{Springer New York}, \bibinfo{year}{1996}).

\bibitem[{\citenamefont{Pargellis et~al.}(1992)\citenamefont{Pargellis, Finn,
  Goodby, Panizza, Yurke, and Cladis}}]{Pargellis}
\bibinfo{author}{\bibfnamefont{A.~N.} \bibnamefont{Pargellis}},
  \bibinfo{author}{\bibfnamefont{P.}~\bibnamefont{Finn}},
  \bibinfo{author}{\bibfnamefont{J.~W.} \bibnamefont{Goodby}},
  \bibinfo{author}{\bibfnamefont{P.}~\bibnamefont{Panizza}},
  \bibinfo{author}{\bibfnamefont{B.}~\bibnamefont{Yurke}}, \bibnamefont{and}
  \bibinfo{author}{\bibfnamefont{P.~E.} \bibnamefont{Cladis}},
  \bibinfo{journal}{Phys. Rev. A} \textbf{\bibinfo{volume}{46}},
  \bibinfo{pages}{7765} (\bibinfo{year}{1992}).

\bibitem[{\citenamefont{Carlsson et~al.}(1995)\citenamefont{Carlsson, Leslie,
  and Clark}}]{Carlsson}
\bibinfo{author}{\bibfnamefont{T.}~\bibnamefont{Carlsson}},
  \bibinfo{author}{\bibfnamefont{F.~M.} \bibnamefont{Leslie}},
  \bibnamefont{and} \bibinfo{author}{\bibfnamefont{N.~A.} \bibnamefont{Clark}},
  \bibinfo{journal}{Phys.\ Rev.\ E} \textbf{\bibinfo{volume}{51}},
  \bibinfo{pages}{4509} (\bibinfo{year}{1995}).

\bibitem[{\citenamefont{Johnson and Saupe}(1977)}]{Johnson}
\bibinfo{author}{\bibfnamefont{D.}~\bibnamefont{Johnson}} \bibnamefont{and}
  \bibinfo{author}{\bibfnamefont{A.}~\bibnamefont{Saupe}},
  \bibinfo{journal}{Phys.\ Rev.\ A} \textbf{\bibinfo{volume}{15}},
  \bibinfo{pages}{2079} (\bibinfo{year}{1977}).

\bibitem[{\citenamefont{Cladis et~al.}(1985)\citenamefont{Cladis, Couder, and
  Brand}}]{Cladis}
\bibinfo{author}{\bibfnamefont{P.~E.} \bibnamefont{Cladis}},
  \bibinfo{author}{\bibfnamefont{Y.}~\bibnamefont{Couder}}, \bibnamefont{and}
  \bibinfo{author}{\bibfnamefont{H.~R.} \bibnamefont{Brand}},
  \bibinfo{journal}{Phys.\ Rev.\ Lett.} \textbf{\bibinfo{volume}{55}},
  \bibinfo{pages}{2945} (\bibinfo{year}{1985}).

\bibitem[{\citenamefont{Chevallard et~al.}(1997)\citenamefont{Chevallard,
  Fisch, and Gilli}}]{Chevallard}
\bibinfo{author}{\bibfnamefont{C.}~\bibnamefont{Chevallard}},
  \bibinfo{author}{\bibfnamefont{T.}~\bibnamefont{Fisch}}, \bibnamefont{and}
  \bibinfo{author}{\bibfnamefont{J.~M.} \bibnamefont{Gilli}},
  \bibinfo{journal}{J. Phys. II (France)} \textbf{\bibinfo{volume}{7}},
  \bibinfo{pages}{1261} (\bibinfo{year}{1997}).

\bibitem[{\citenamefont{Clark}(1979)}]{ClarkSmC}
\bibinfo{author}{\bibfnamefont{N.}~\bibnamefont{Clark}},
  \bibinfo{journal}{Appl. Phys. Lett.} \textbf{\bibinfo{volume}{35}},
  \bibinfo{pages}{688} (\bibinfo{year}{1979}).

\bibitem[{\citenamefont{Clark and Meyer}(1973)}]{ClarkSmA}
\bibinfo{author}{\bibfnamefont{N.}~\bibnamefont{Clark}} \bibnamefont{and}
  \bibinfo{author}{\bibfnamefont{R.~B.} \bibnamefont{Meyer}},
  \bibinfo{journal}{Appl. Phys. Lett.} \textbf{\bibinfo{volume}{22}},
  \bibinfo{pages}{493} (\bibinfo{year}{1973}).

\bibitem[{\citenamefont{Yablonskii et~al.}(2003)\citenamefont{Yablonskii,
  Nakano, Ozaki, and Yoshino}}]{Yablonskii}
\bibinfo{author}{\bibfnamefont{S.~V.} \bibnamefont{Yablonskii}},
  \bibinfo{author}{\bibfnamefont{K.}~\bibnamefont{Nakano}},
  \bibinfo{author}{\bibfnamefont{M.}~\bibnamefont{Ozaki}}, \bibnamefont{and}
  \bibinfo{author}{\bibfnamefont{K.}~\bibnamefont{Yoshino}},
  \bibinfo{journal}{JETP Lett.} \textbf{\bibinfo{volume}{77}},
  \bibinfo{pages}{140} (\bibinfo{year}{2003}).

\bibitem[{\citenamefont{Uto et~al.}(1997)\citenamefont{Uto, Tazoh, Ozaki, and
  Yoshino}}]{Uto}
\bibinfo{author}{\bibfnamefont{S.}~\bibnamefont{Uto}},
  \bibinfo{author}{\bibfnamefont{E.}~\bibnamefont{Tazoh}},
  \bibinfo{author}{\bibfnamefont{M.}~\bibnamefont{Ozaki}}, \bibnamefont{and}
  \bibinfo{author}{\bibfnamefont{K.}~\bibnamefont{Yoshino}},
  \bibinfo{journal}{J.\ Appl.\ Phys.} \textbf{\bibinfo{volume}{82}},
  \bibinfo{pages}{2791} (\bibinfo{year}{1997}).

\bibitem[{\citenamefont{Luk'yanchuk}(1998)}]{Luk'yanchuk}
\bibinfo{author}{\bibfnamefont{I.}~\bibnamefont{Luk'yanchuk}},
  \bibinfo{journal}{Phys.\ Rev.\ E} \textbf{\bibinfo{volume}{57}},
  \bibinfo{pages}{574} (\bibinfo{year}{1998}).

\bibitem[{\citenamefont{Kundagrami and Lubensky}(2003)}]{Kundagrami}
\bibinfo{author}{\bibfnamefont{A.}~\bibnamefont{Kundagrami}} \bibnamefont{and}
  \bibinfo{author}{\bibfnamefont{T.~C.} \bibnamefont{Lubensky}},
  \bibinfo{journal}{Phys.\ Rev.\ E} \textbf{\bibinfo{volume}{68}},
  \bibinfo{pages}{060703} (\bibinfo{year}{2003}).

\bibitem[{\citenamefont{Navailles et~al.}(1995)\citenamefont{Navailles, Pindak,
  Barois, and Nguyen}}]{NavaillesTGBC}
\bibinfo{author}{\bibfnamefont{L.}~\bibnamefont{Navailles}},
  \bibinfo{author}{\bibfnamefont{R.}~\bibnamefont{Pindak}},
  \bibinfo{author}{\bibfnamefont{P.}~\bibnamefont{Barois}}, \bibnamefont{and}
  \bibinfo{author}{\bibfnamefont{H.~T.} \bibnamefont{Nguyen}},
  \bibinfo{journal}{Phys.\ Rev.\ Lett.} \textbf{\bibinfo{volume}{74}},
  \bibinfo{pages}{5224} (\bibinfo{year}{1995}).

\bibitem[{\citenamefont{Meiboom and Hewitt}(1975)}]{Meiboom}
\bibinfo{author}{\bibfnamefont{A.}~\bibnamefont{Meiboom}} \bibnamefont{and}
  \bibinfo{author}{\bibfnamefont{R.~C.} \bibnamefont{Hewitt}},
  \bibinfo{journal}{Phys.\ Rev.\ Lett.} \textbf{\bibinfo{volume}{34}},
  \bibinfo{pages}{1146} (\bibinfo{year}{1975}).

\bibitem[{\citenamefont{Chen et~al.}(1990)\citenamefont{Chen, Jayaprakash,
  Pandit, and Wenzel}}]{ChenJayaprakash}
\bibinfo{author}{\bibfnamefont{K.}~\bibnamefont{Chen}},
  \bibinfo{author}{\bibfnamefont{C.}~\bibnamefont{Jayaprakash}},
  \bibinfo{author}{\bibfnamefont{R.}~\bibnamefont{Pandit}}, \bibnamefont{and}
  \bibinfo{author}{\bibfnamefont{W.}~\bibnamefont{Wenzel}},
  \bibinfo{journal}{Phys.\ Rev.\ Lett.} \textbf{\bibinfo{volume}{65}},
  \bibinfo{pages}{2736} (\bibinfo{year}{1990}).

\bibitem[{\citenamefont{Safran et~al.}(1986)\citenamefont{Safran, Robbins, and
  Garoff}}]{SafranRobbins}
\bibinfo{author}{\bibfnamefont{S.~A.} \bibnamefont{Safran}},
  \bibinfo{author}{\bibfnamefont{M.~O.} \bibnamefont{Robbins}},
  \bibnamefont{and} \bibinfo{author}{\bibfnamefont{S.}~\bibnamefont{Garoff}},
  \bibinfo{journal}{Phys.\ Rev.\ A} \textbf{\bibinfo{volume}{33}},
  \bibinfo{pages}{2186} (\bibinfo{year}{1986}).

\bibitem[{\citenamefont{Wang and Gong}(1996)}]{WangGong}
\bibinfo{author}{\bibfnamefont{Z.}~\bibnamefont{Wang}} \bibnamefont{and}
  \bibinfo{author}{\bibfnamefont{C.}~\bibnamefont{Gong}},
  \bibinfo{journal}{Phys.\ Rev.\ B} \textbf{\bibinfo{volume}{54}},
  \bibinfo{pages}{17067} (\bibinfo{year}{1996}).

\bibitem[{\citenamefont{Kuhn and Rehage}(2000)}]{KuhnRehage}
\bibinfo{author}{\bibfnamefont{H.}~\bibnamefont{Kuhn}} \bibnamefont{and}
  \bibinfo{author}{\bibfnamefont{H.}~\bibnamefont{Rehage}},
  \bibinfo{journal}{Phys.\ Chem.\ Chem.\ Phys.} \textbf{\bibinfo{volume}{2}},
  \bibinfo{pages}{1023} (\bibinfo{year}{2000}).

\bibitem[{\citenamefont{Tamamushi}(1974)}]{Tamamushi}
\bibinfo{author}{\bibfnamefont{B.}~\bibnamefont{Tamamushi}},
  \bibinfo{journal}{Rheol.\ Acta} \textbf{\bibinfo{volume}{13}},
  \bibinfo{pages}{247} (\bibinfo{year}{1974}).

\bibitem[{\citenamefont{Pieranski et~al.}(1993)\citenamefont{Pieranski,
  Beliard, Tournellec, Leoncini, Furtlehner, Dumoulin, Riou, Jouvin,
  F{\'e}nerol, Palaric et~al.}}]{Pieranski}
\bibinfo{author}{\bibfnamefont{P.}~\bibnamefont{Pieranski}},
  \bibinfo{author}{\bibfnamefont{L.}~\bibnamefont{Beliard}},
  \bibinfo{author}{\bibfnamefont{J.-P.} \bibnamefont{Tournellec}},
  \bibinfo{author}{\bibfnamefont{X.}~\bibnamefont{Leoncini}},
  \bibinfo{author}{\bibfnamefont{C.}~\bibnamefont{Furtlehner}},
  \bibinfo{author}{\bibfnamefont{H.}~\bibnamefont{Dumoulin}},
  \bibinfo{author}{\bibfnamefont{E.}~\bibnamefont{Riou}},
  \bibinfo{author}{\bibfnamefont{B.}~\bibnamefont{Jouvin}},
  \bibinfo{author}{\bibfnamefont{J.-P.} \bibnamefont{F{\'e}nerol}},
  \bibinfo{author}{\bibfnamefont{P.}~\bibnamefont{Palaric}},
  \bibnamefont{et~al.}, \bibinfo{journal}{Physica A}
  \textbf{\bibinfo{volume}{194}}, \bibinfo{pages}{364} (\bibinfo{year}{1993}).

\bibitem[{\citenamefont{Chen and Jasnow}(2000)}]{ChenJasnow}
\bibinfo{author}{\bibfnamefont{H.-Y.} \bibnamefont{Chen}} \bibnamefont{and}
  \bibinfo{author}{\bibfnamefont{D.}~\bibnamefont{Jasnow}},
  \bibinfo{journal}{Phys.\ Rev.\ E} \textbf{\bibinfo{volume}{61}},
  \bibinfo{pages}{493} (\bibinfo{year}{2000}).

\bibitem[{\citenamefont{Galerne}(1981)}]{Galerne}
\bibinfo{author}{\bibfnamefont{Y.}~\bibnamefont{Galerne}},
  \bibinfo{journal}{Phys.\ Rev.\ A} \textbf{\bibinfo{volume}{24}},
  \bibinfo{pages}{2284} (\bibinfo{year}{1981}).

\bibitem[{\citenamefont{Kuo et~al.}(2006)\citenamefont{Kuo, Lee, and Wu}}]{Kuo}
\bibinfo{author}{\bibfnamefont{L.-Y.} \bibnamefont{Kuo}},
  \bibinfo{author}{\bibfnamefont{K.-L.} \bibnamefont{Lee}}, \bibnamefont{and}
  \bibinfo{author}{\bibfnamefont{J.-J.} \bibnamefont{Wu}},
  \bibinfo{journal}{Jpn.\ J.\ Appl.\ Phys.} \textbf{\bibinfo{volume}{45}},
  \bibinfo{pages}{8775} (\bibinfo{year}{2006}).

\end{thebibliography}

\end{document}